\documentclass[useAMS,usenatbib,usegraphicx]{mn2e}
\usepackage{graphicx}                                            
\usepackage{times}
\usepackage{float}
\usepackage{rotating}
\usepackage{epstopdf}
\usepackage{multirow}
\usepackage{pdflscape}
\usepackage{color}
\usepackage{tabularx}

\newcommand{\ergps}{erg~s$^{-1}$}

\def\ltsima{$\; \buildrel < \over \sim \;$}
\def\simlt{\lower.5ex\hbox{\ltsima}}
\def\gtsima{$\; \buildrel > \over \sim \;$}
\def\simgt{\lower.5ex\hbox{\gtsima}}
\def\gsimeq
{\hbox{\raise0.5ex\hbox{$>\lower1.06ex\hbox{$\kern-1.07em{\sim}$}$}}}
\def\lsimeq
{\hbox{\raise0.5ex\hbox{$<\lower1.06ex\hbox{$\kern-1.07em{\sim}$}$}}}

\def\xmm{{\it XMM-Newton }}

\def\xmm{{\it XMM-Newton}}
\def\chandra{{\it Chandra}}

\def\nustar{{\it NuSTAR}}
\def\swift{{\it Swift}}

\def\apj{ApJ}
\def\mnras{MNRAS}
\def\aap{A\&A}
\def\apjl{ApJ}
\def\apjs{ApJS}

\def\pasj{PASJ}
\def\nat{Nature}
\def\iaucirc{IAU~circular}

\def\ssr{SSRv}

\def\exo{EXO~0748-676}
\def\axj{AX~J1745.6-2901}

\def\mxb{MXB~1659-298}

\def\xspec{{\it Xspec}}

\def\Fetwt{Fe~{\sc xxii}}
\def\Fevc{Fe~{\sc xxv}}
\def\Fevs{Fe~{\sc xxvi}}

\def\xis{XIS}
\def\xis1{XIS1}
\def\xis2{XIS2}
\def\xis3{XIS3}

\title[] 
 {{Evolution of the disc atmosphere in the X-ray binary \mxb, 
 during its 2015-2017 outburst}} 
 \author[G.\ Ponti et al. ]
 {G.~Ponti$^{1,2}$, S. Bianchi$^{3}$, B. De Marco$^{4}$, 
 A. Bahramian$^{5,6}$, N. Degenaar$^{7}$ and C.~O. Heinke$^{8}$ 
 \\
   $^1$ INAF-Osservatorio Astronomico di Brera, Via E. Bianchi 46, I-23807 Merate (LC), Italy\\
   $^2$ Max-Planck-Institut f{\"u}r Extraterrestrische Physik, Giessenbachstrasse, D-85748, Garching, Germany\\
   $^3$ Dipartimento di Matematica e Fisica, Universit\`a Roma Tre, Via della Vasca Navale 84, I-00146, Roma, Italy\\
   $^4$ Nicolaus Copernicus Astronomical Center, PL-00-716 Warsaw, Poland\\
   $^5$ Department of Physics and Astronomy, Michigan State University, East Lansing, MI 48824\\
   $^6$ International Centre for Radio Astronomy Research Ð Curtin University, GPO Box U1987, Perth, WA 6845, Australia\\
   $^7$ Anton Pannekoek Institute for Astronomy, University of Amsterdam, Science Park 904, 1098 XH Amsterdam, The Netherlands\\
   $^8$ Department of Physics, CCIS 4-183, University of Alberta, Edmonton, AB T6G 2E1, Canada\\
      }
\pagerange{\pageref{firstpage}--\pageref{lastpage}}
\usepackage{times}
\begin{document}
\label{firstpage}
 \maketitle
\begin{abstract}
We report on the evolution of the X-ray emission of the accreting 
neutron star (NS) low mass X-ray binary (LMXB), \mxb, during 
its most recent outburst in 2015-2017. We detected 60 absorption 
lines during the soft state (of which 21 at more than $3 \sigma$), 
that disappeared in the hard state (e.g., the \Fevc\ and \Fevs\ lines). 
The absorbing plasma is at rest, likely part of the accretion disc 
atmosphere. The bulk of the absorption features can be reproduced 
by a high column density ($log(N_H/cm^{-2})\sim23.5$) 
of highly ionised ($log(\xi/erg~cm~s^{-1})\sim3.8$) 
plasma. Its disappearance during the hard state 
is likely the consequence of a thermal photo-ionisation instability. 
\mxb's continuum emission can be described by the sum of 
an absorbed disk black body and its Comptonised emission, plus a black body 
component. The observed spectral evolution with state is in line 
with that typically observed in atoll and stellar mass black hole 
LMXB. The presence of a relativistic Fe~K$\alpha$ disk-line 
is required during the soft state. We also tentatively detect 
the \Fetwt\ doublet, whose ratio suggests an electron density 
of the absorber of $n_e>10^{13}$~cm$^{-3}$, hence, the absorber 
is likely located at $<7\times10^4$~r$_g$ from the illuminating source, 
well inside the Compton and outer disc radii. 
\mxb\ is the third well monitored atoll LMXB showcasing 
intense \Fevc\ and \Fevs\ absorption during the soft state 
that disappears during the hard state. 
\end{abstract}


\section{Introduction}

Winds are fundamental components of accretion onto 
X-ray binaries (Shakura \& Sunyaev 1973; Begelman et al. 1983). 
In black hole (BH) X-ray binaries, they are observed to have 
an equatorial geometry and a strong state dependence (Miller et al. 2006; 
Neilsen \& Lee 2009; Ponti et al. 2012; Neilsen et al. 2018). 
Winds might even affect the effective viscosity of the accretion disc, 
as well as the X-ray binary (orbital period) evolution (Ponti et al. 2017; 
Tetarenko et al. 2018). 

Analogously to black hole (BH), accreting neutron star (NS) systems 
also show ionised absorbers with similar equatorial geometry, 
similar column densities and ionisations (Ueda et al. 2004; 
Diaz Trigo et al. 2006; 2016; Ponti et al. 2016). 
It is still an open question whether the highly ionised absorbers 
(traced by \Fevc\ and \Fevs) in NS share the same state dependence 
as the ones observed in accreting BH. 
Recent investigations focussed on \exo\ and \axj, which are two of the 
atoll NS systems with the best monitoring in the Fe~K band, covering 
both soft and hard states. In both sources it was observed that 
the highly ionised absorption is stronger during the soft state, 
while it is undetected during the hard state (Ponti et al. 2014; 2015; 2018). 
This might suggest that the state-absorption connection is a common 
property of NS (atoll) sources. 

Although the origin of such state-absorption connection is still debated, 
the results of photo-ionisation computations clearly showed 
that whenever the soft state absorber in X-ray binaries is illuminated 
by the hard state spectral energy distribution (SED), the plasma becomes unstable 
(Jimenez-Garate et al. 2001; Chakravorty et al. 2013; 
Higgibottom \& Proga 2015; Bianchi et al. 2017). As a result of such photo-ionisation 
thermal instability the plasma is prone to change its properties (e.g., condensing, 
expanding, etc.) and it will likely migrate to other stable solutions 
(Bianchi et al. 2017).  

To further characterise the properties of the absorbers in accreting NS, 
as well as to further investigate the state-absorption connection in atoll sources, 
we triggered \xmm, \chandra\ and \nustar\ observations during the last 
($\sim1.5$ years long) outburst of \mxb, that started in August 2015 
(Negoro et al. 2015; Bahramian et al. 2015). 

\mxb\ is a transient atoll low mass X-ray binary displaying type-I X-ray bursts, 
therefore indicating a neutron star primary (Lewin et al. 1976; Galloway et al. 2008). 
\mxb\ is a high inclination system ($i\sim73-78^\circ$; Frank et al. 1987; 
Ponti et al. 2018), showing dipping and eclipsing events, with an orbital period of $P_{orb}=7.1$~hr 
(Cominski \& Wood 1984; 1989; Jain et al. 2017; Iaria et al. 2018). 
The previous outburst of \mxb\ started on April 1999 and lasted 
for $\sim2.5$ years (Wijnands et al. 2002). During the 1999-2001 outburst, \xmm\ 
observed \mxb\ twice, showing clear evidence for \Fevc\ and \Fevs, as well as lower 
ionisation, absorption lines (Sidoli et al. 2001), however, no \chandra\ HETG 
observation was performed, to best detail the properties of the absorber 
in the Fe~K band. 

We report here the analysis of the \xmm, \chandra, \nustar\ and 
\swift\ monitoring campaigns of the last outburst of \mxb.

\section{Assumptions}

All spectral fits were performed using the \xspec\ software package 
(version 12.7.0; Arnaud 1996). 
Uncertainties and upper limits are reported at the 90 per cent confidence 
level for one interesting parameter, unless otherwise stated. 
Using X-ray bursts, a distance to \mxb\ of $9\pm2$ or $12\pm3$~kpc has 
been inferred by Galloway et al. (2008) for a hydrogen or helium-rich 
ignition layer, respectively. We noted that \mxb\ appears in Gaia DR2, 
however without parallax estimate, hence consistent with the large distance 
suggested by previous estimates (Gaia Collaboration et al. 2018). 
All luminosities, black body and disc black body radii assume 
that \mxb\ is located at 10 kpc. To derive 
the disc black body inner radius $r_{DBB}$, we fit the spectrum with the 
{\sc diskbb} model in {\sc XSpec} (Mitsuda et al. 1984; Makishima et al. 1986)
the normalisation of which provides the apparent inner disc radius ($R_{DBB}$). 
Following Kubota et al. (1998), we correct the apparent inner 
disc radius through the equation: $r_{DBB}=\xi \kappa^2 R_{DBB}$ 
(where $\kappa=2$ and $\xi=\sqrt{(3/7)}\times(6/7)^3$) in order to estimate 
the real inner disc radius $r_{DBB}$. We also assume an inclination of the 
accretion disc of $75^{\circ}$ (Frank et al. 1987; Ponti et al. 2018). 
We adopt a nominal Eddington limit 
for \mxb\ of  L$_{Edd}=2\times10^{38}$~\ergps (appropriate for a primary mass 
of $M_{NS}\sim1.4$~M$_{\odot}$ and cosmic composition; Lewin et al. 1993). 
We use the $\chi^2$ statistics to fit CCD resolution spectra (we group each 
spectrum to have a minimum of 30 counts in each bin), while we employ Cash 
statistics (Cash 1979) to fit the high resolution un-binned ones. 
We fit the interstellar absorption 
with the {\sc tbabs} model in \xspec\ assuming Wilms et al. (2000) abundances 
and Verner et al. (1996) cross sections. 

\section{Observations and data reduction}

At the beginning of the 2015-2017 outburst of \mxb, we requested 
a 40~ks \xmm\ observation, to either discover or rule out 
the presence of Fe~K absorption during the hard state of this source 
(archival observations only caught the soft state). 
\xmm\ observed \mxb\ on 2015-09-26 (obsid 0748391601). 
All EPIC cameras were in timing mode with the thin filter applied. 
The data were analysed with the latest version (17.0.0) of the \xmm\ 
(Jansen et al. 2001; Str\"{u}der et al. 2001; Turner et al. 2001) 
Science Analysis System {\sc sas}, applying the most recent 
(as of 2017 September 20) calibrations. 
We reduced the data with the standard pipelines ({\sc epchain}, 
{\sc emchain} and {\sc rgsproc} for the EPIC-pn, EPIC-MOS and 
RGS camera, respectively). 
Because of the higher effective area, we show EPIC-pn data only, 
in addition to the RGS.  
We extracted the EPIC-pn source photons within {\sc rawx} 20 and 53 
(while the background between 2 and 18), {\sc pattern$<$=4} 
and {\sc flag==0}. 

Following a softening of the X-ray emission, we triggered pre-approved 
\chandra\ observations (performed on 2016 April $21^{\rm st}$) 
of \mxb, to detail the ionised absorber properties during the soft state.  
The \chandra\ spectra and response matrices have been produced with 
the {\sc chandra\_repro} task, reducing the width of the default spatial mask 
to avoid overlap of the HEG and MEG boxes at high energy ($E\geq7.3$~keV). 

\nustar\ (Harrison et al. 2013) observed \mxb\ twice during its 2015-2017 
outburst, on 2015 September $28^{\rm th}$ (obsid 90101013002), two days 
after the \xmm\ observation and on 2016 April $21^{\rm st}$ (90201017002), 
simultaneous with the \chandra\ one. 
The \nustar\ data were reduced with the standard {\it nupipeline} 
scripts v. 0.4.5 (released on 2016-03-30) and the high level products produced 
with the {\it nuproducts} tool. 
The source and background photons were extracted from circular regions 
of $120^{\prime\prime}$ and $180^{\prime\prime}$ radii, respectively, 
centred on the source and at the corner of the CCD, respectively. 
Response matrices appropriate for each data-set were generated using 
the standard software. We did not combine modules. 
Bursts, dips and eclipses have been removed by generating a light curve 
in {\sc xselect} with 60~s time binning in the 3-70 keV energy band and cutting 
all intervals with a count rate outside the persistent value (e.g., 1.7-3.3 and 
10-11 ph~s$^{-1}$ for the two observations). 

The outburst of \mxb\ was monitored with the Neil Gehrels {\it Swift} 
Observatory, to characterise the long term evolution of the 
spectral energy distribution (SED). Full details on the analysis of the \swift\ data 
will be presented in a forthcoming paper (Degenaar et al. in prep). 
We analysed the Ultra-Violet/Optical Telescope (Roming et al. 2005) data 
obtained on the same day of the \xmm\ (obsid: 00034002012; 
filter {\sc w1}), the first \nustar\ (00081770001; {\sc v,b,u,w1,w2,m2}) 
and the \chandra+\nustar\ (00081918001; {\sc v,b,u,w1,w2,m2}) 
observations. The various filters employed allowed us to cover 
a wavelength range of $\simeq 1500-8500$~\AA{} (Poole et al. 2008). 
To extract source photons we used a standard aperture of $3^\prime$, 
whereas for the background we used a source-free region 
with a radius of $9^\prime$. For each observation/filter we first combined 
all image extensions using \textsc{uvotimsum} and then extracted 
magnitudes and fluxes using \textsc{uvotsource}. 

\section{Timing analysis and state determination}
\label{timing}

To determine the evolution of the state of the source between the various 
observations, we first extracted light curves in the 3-10 keV energy range, 
with time bins of 6~ms\footnote{The \chandra\ HETG 
frame time does not allow to constrain the spectral state through timing. }. 
We computed the fractional rms in the range 0.002-64~Hz for the \xmm\ and 
\nustar\ observations. From the \xmm\ observation we estimated a fractional 
rms of $21\pm4$~\%, this value being typical of the hard state 
(Mu\~noz-Darias et al. 2011, 2014). 
In the case of \nustar\ observations, we followed the method described 
in Bachetti et al. (2015) in order to properly account for dead time effects. 
Given the poorer statistics (due to the use of the co-spectrum between 
the light curves of the two modules) only an upper limit to the 
0.002-64~Hz fractional rms of $< 8$~\% can be derived from the 
second \nustar\ observation, suggesting that \mxb\ was in the soft 
state at that time. On the other hand, the rms could not be constrained 
from the first \nustar\ observation. 

\begin{figure}
\vspace{-0.15cm}

\hspace{-0.5cm}
\includegraphics[height=0.4\textwidth,angle=-0]{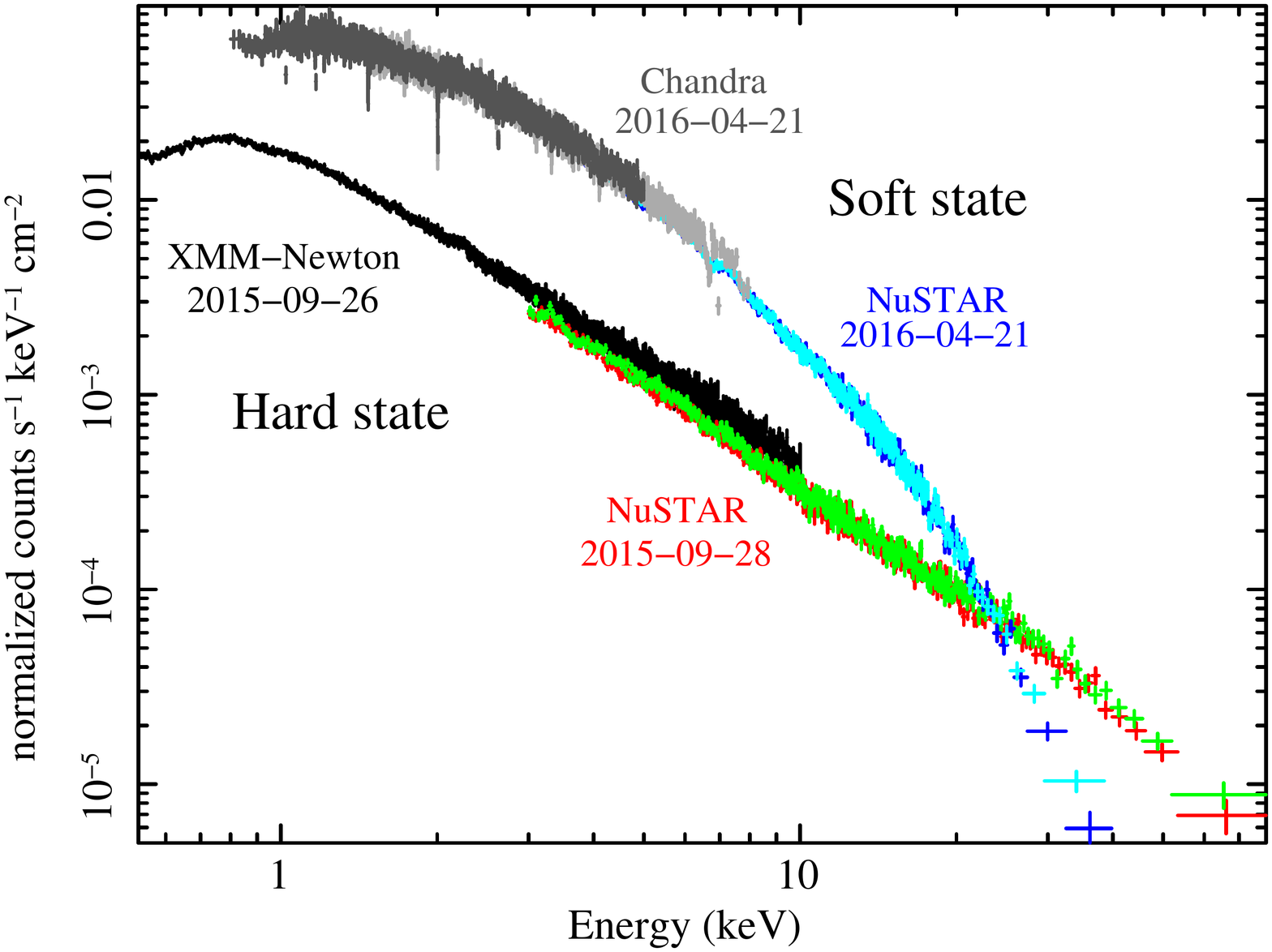}
\caption{\xmm, \nustar\  and \chandra\ spectra of \mxb\ during the 2015-2017 outburst. 
Both the \xmm\ (black) and the \nustar\ (red and green) spectra accumulated 
on 2015-09-26 and 2015-09-28, respectively, show a hard power law 
shape characteristic of the hard state. 
The simultaneous \chandra\ (light and dark grey) and the \nustar\ (light and 
dark blue) spectra obtained in 2016-04-21 show a significantly softer and 
brighter emission below $\sim20$~keV, with a significant drop 
above $\sim20-30$~keV. }
\label{spect}
\end{figure}

Figure \ref{spect} shows the \xmm, \nustar\ and \chandra\ spectra of \mxb\ 
during the 2015-2017 outburst. In agreement with the timing results, the \xmm\ 
spectrum (black data) is characterised by a hard power 
law shape, confirming that \mxb\ was in the hard state at that time. 
The first \nustar\ dataset (red and green data), accumulated two days after, 
also shows a very similar hard spectrum, suggestive of a hard state (Fig. \ref{spect}). 
On the other hand, during the simultaneous \chandra\ (dark and light grey) 
and second \nustar\ (dark and light blue) observations, \mxb\ displays a softer 
and significantly brighter emission, below $\sim20$~keV, while the source emission 
drops quickly above $\sim20-30$~keV. In agreement with the timing results, 
this indicates that the source was in the soft state during these observations. 

\section{Broad band fit of the soft state with approximate models}
\label{BBfit}

\begin{figure}
\vspace{-0.3cm}

\hspace{-0.5cm}
\vspace{-0.5cm}
\includegraphics[height=0.4\textwidth,angle=0]{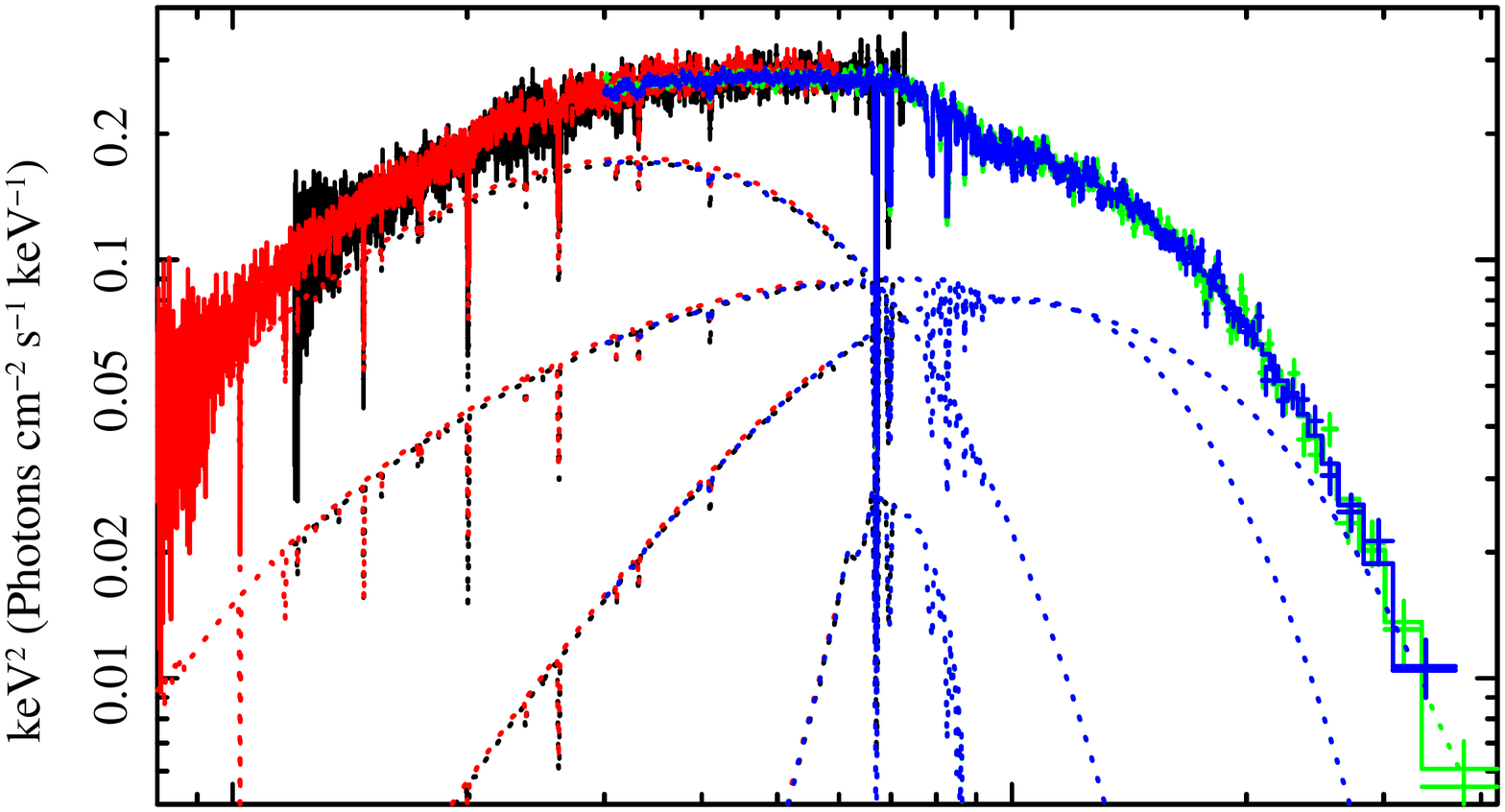}

\vspace{-5.5cm}
\hspace{-0.5cm}
\includegraphics[height=0.4\textwidth,angle=0]{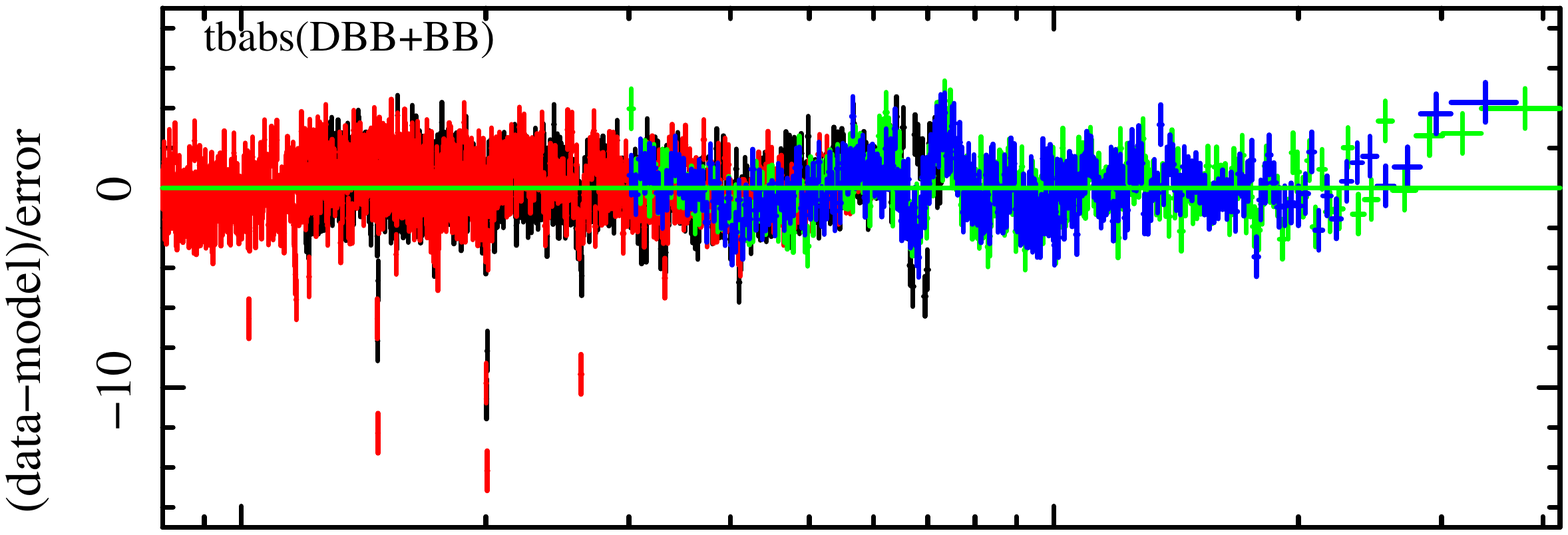}

\vspace{-4.5cm}
\hspace{-0.5cm}
\includegraphics[height=0.4\textwidth,angle=0]{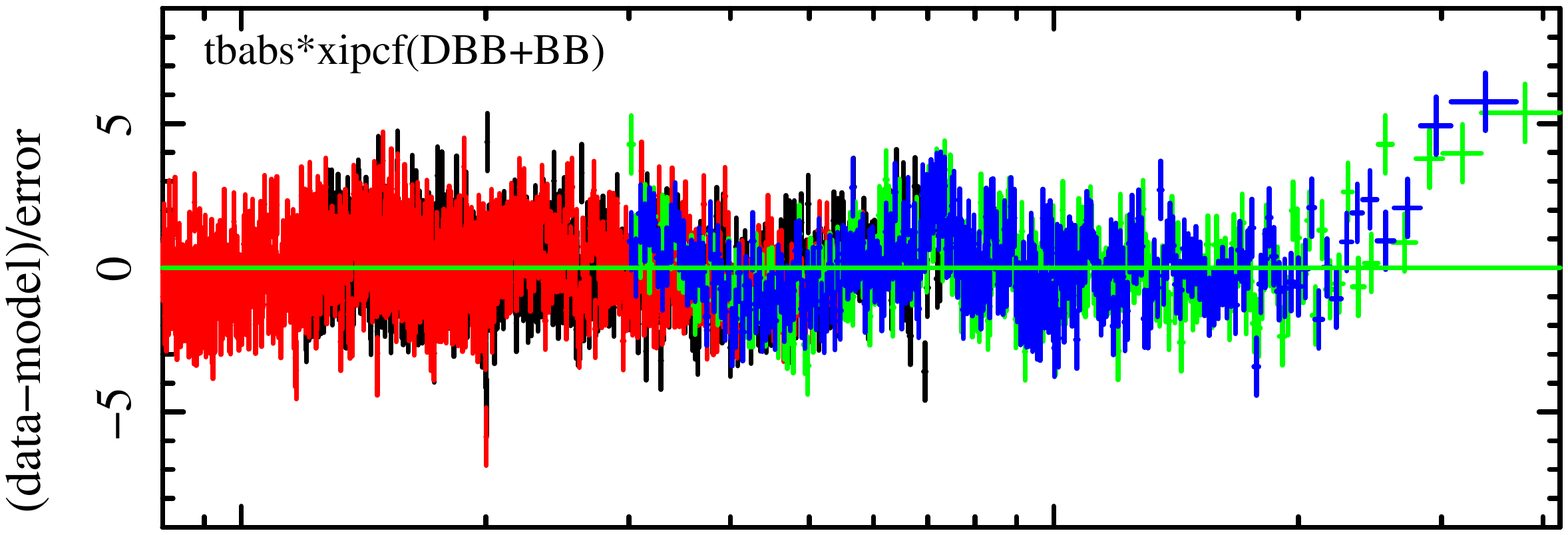}

\vspace{-4.5cm}
\hspace{-0.5cm}
\includegraphics[height=0.4\textwidth,angle=0]{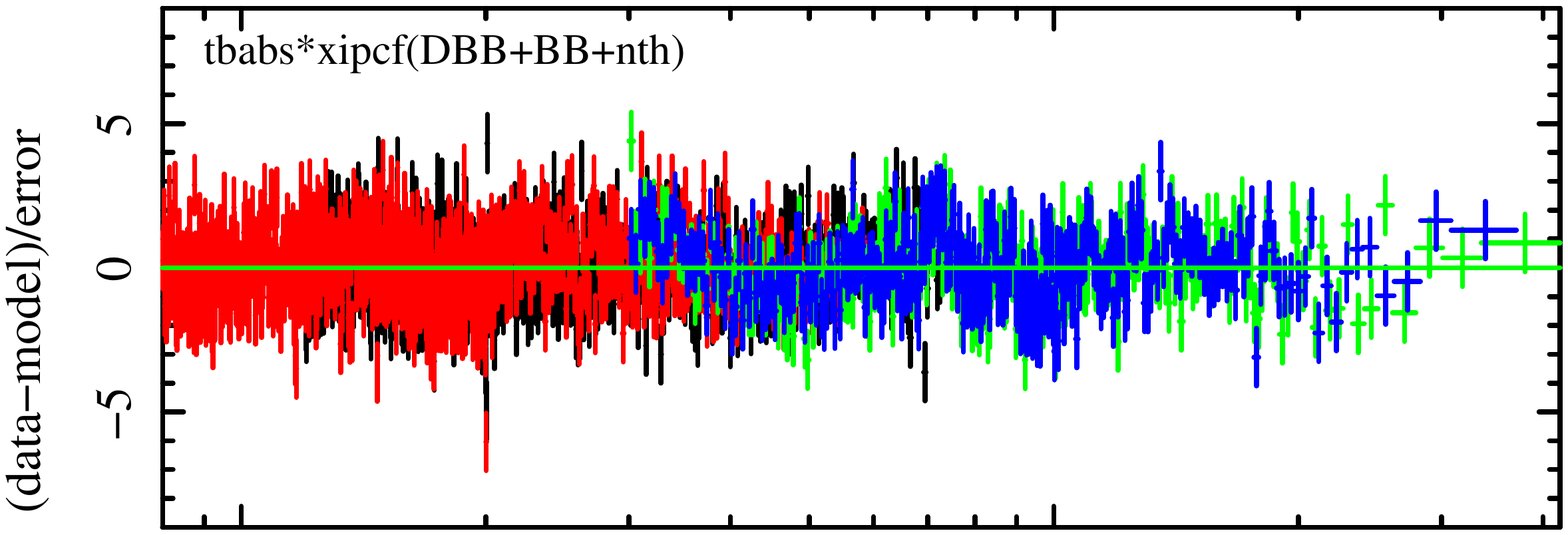}

\vspace{-4.5cm}
\hspace{-0.5cm}
\includegraphics[height=0.4\textwidth,angle=0]{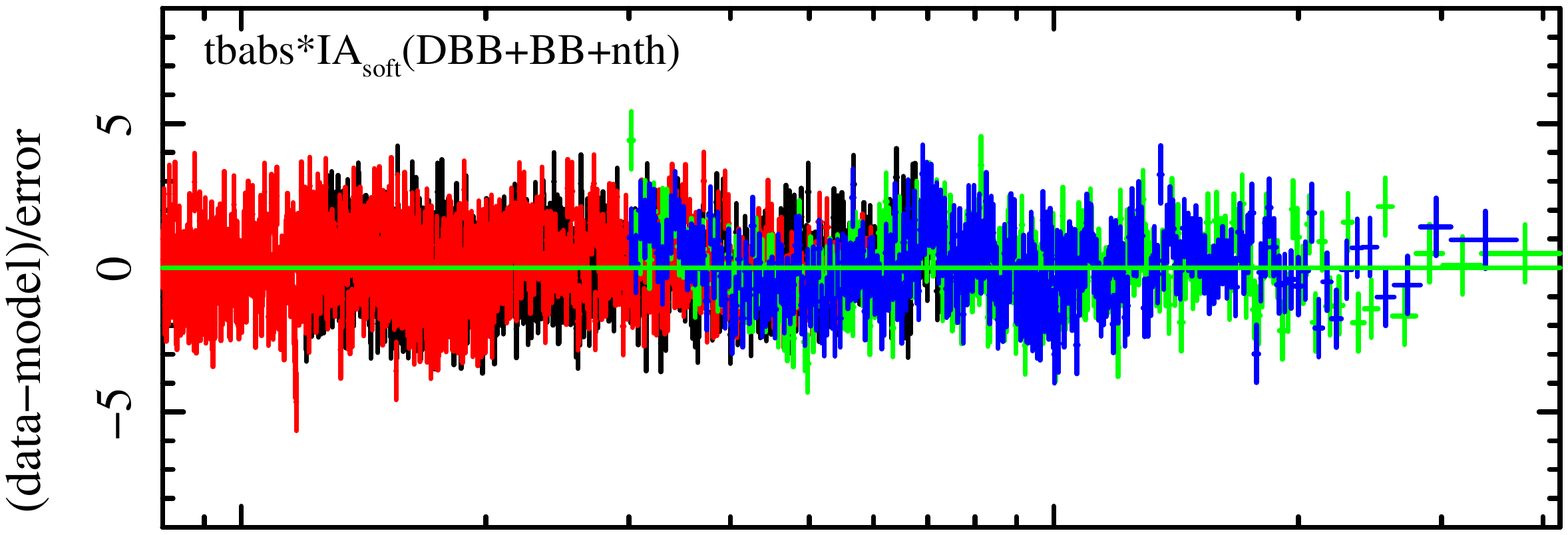}

\vspace{-4.5cm}
\hspace{-0.5cm}
\includegraphics[height=0.4\textwidth,angle=0]{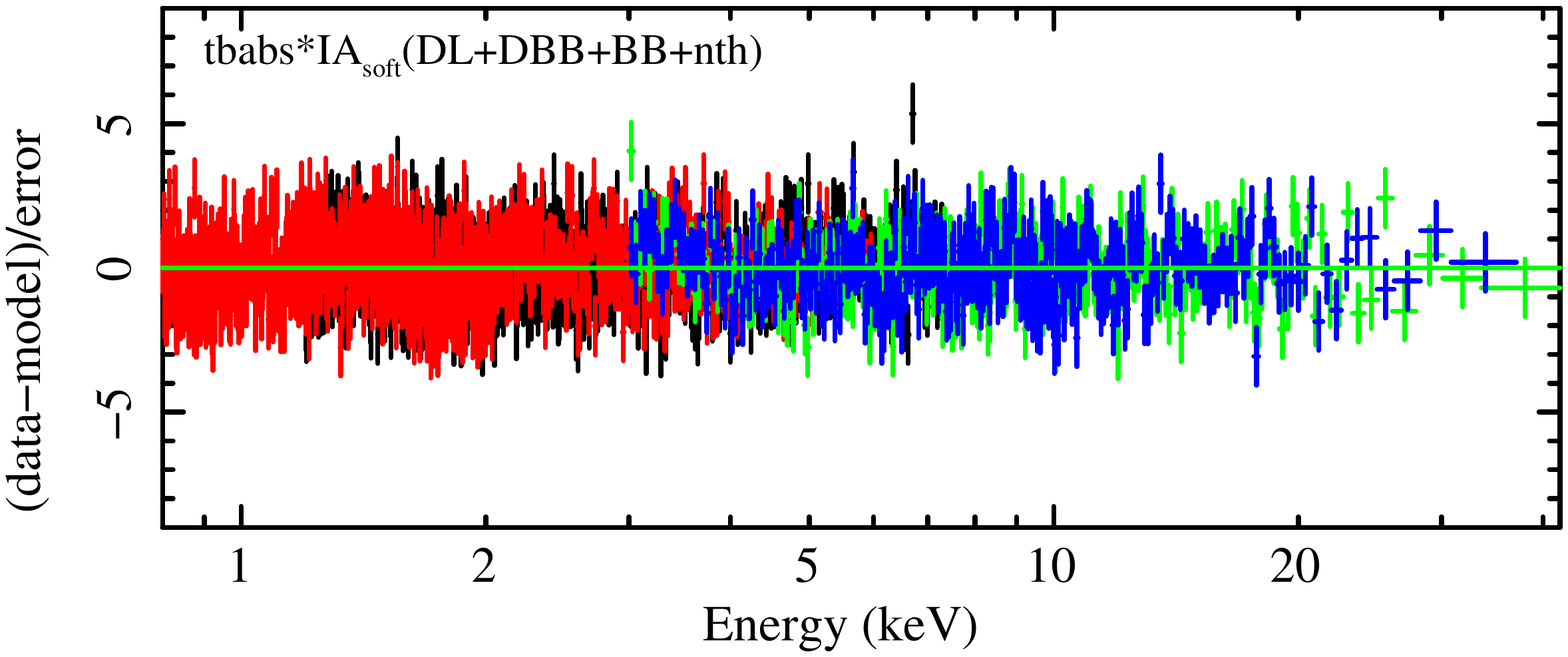}
\caption{Soft state persistent emission of \mxb\ fitted with phenomenological models. 
MEG, HEG, FPMA and FPMB spectra and residuals 
are shown in red, black, green and blue, respectively. 
{\it (Top panel) } Best fit of the soft state spectrum fitted with disk black-body plus 
black-body plus Comptonisation plus a disk-line emission absorbed by neutral 
and ionised material (Tab. \ref{Nutab}). 
From top to bottom the various panels show the residuals for increasingly 
complex models. The first plot shows the residuals of the fit with a disk black 
body plus black body model absorbed by neutral material. Then the 
residuals after the addition of an approximated ionised absorption 
{\sc zxipcf} component. Then, residuals after the inclusion of a thermal 
Comptonisation component ({\sc nthcomp}). Subsequently, 
residuals after the inclusion of a proper ionised absorption component 
({\sc $IA_{soft}$}). Finally, residuals after the inclusion of an additional 
relativistic Fe~K emission line. }
\label{NuHETGSPEC}
\end{figure}
We started by simultaneously fitting the \chandra\ HETG (HEG and MEG first 
order) + \nustar\ data, leaving the cross-normalisation constants ($c_{MEG}$, 
$c_{NuA}$ and $c_{NuB}$) free to vary (see Fig. \ref{NuHETGSPEC}; Tab. \ref{Nutab}). 
We fit the MEG, HEG and FPM(A,B) spectra in the 0.8-6, 1.2-7.3 and 
3-45~keV ranges, respectively. 

The black, red, green and blue points in Fig. \ref{NuHETGSPEC} show 
the \chandra\ HEG and MEG, the \nustar\ FPMA and FPMB spectra 
of the persistent emission, accumulated during the soft state observations 
(dips, eclipses and bursts have been removed; see black data in Fig. 2 of Ponti et al. 2018). 
The fit of these spectra with a model composed by a disk blackbody plus 
a hot blackbody (B1 in Tab. \ref{Nutab}), both absorbed by neutral 
material ({\sc tbabs(diskbb+bbody)} in {\sc XSpec}), provided a reasonable 
description of the continuum, however very significant residuals appeared 
as clear signatures of ionised absorption lines in the soft X-ray 
and Fe~K band (Fig. \ref{NuHETGSPEC}), 
making this model unacceptable ($C-stat=9777.8$ for 7094 dof, Tab. \ref{Nutab}). 
We also noted an excess of emission at high energy 
($E>25$~keV; Fig. \ref{NuHETGSPEC}), that does not appear to be an 
artefact of the background (contributing at $E\geq35$~keV).
\begin{table*}
\begin{center}
\begin{tabular}{ l l c c c c c c c c c c c c c c c c}
\hline
\hline
\multicolumn{7}{c}{\bf \chandra+\nustar\ soft state (2016-04-21)} \\
\hline
Model           &            & {\sc B1}        & {\sc B2}            & {\sc B3}               & {\sc B6}             &{\sc B7}                  \\
                &            & {\sc DBB+BB}    & {\sc xipcf(DBB+BB)} & {\sc xipcf(DBB+BB+nth)}& {\sc IA(DBB+BB+nth)} &{\sc IA(dl+DBB+BB+nth)}   \\
\hline				
$N_{H}$         &$\star$     & $0.156\pm0.008$ & $0.154\pm0.009$     & $0.229\pm0.016$        & $0.233\pm0.007$      & $0.21\pm0.01$            \\
$log(\xi_{1})$  &            &                 & $4.25\pm0.02$       & $4.25\pm0.02$          & $4.54\pm0.01$        & $3.76^{+0.12}_{-0.03}$   \\
$log(N_{H_{1}})$&$\star$     &                 & $23.6\pm0.1$        & $23.6\pm0.1$           & $24.25\pm0.02$       & $23.48^{+0.16}_{-0.05}$  \\
$\alpha_d$      &            &                 &                     &                        &                      & $-2.61^{+0.07}_{-0.10}$  \\ 
$N_d$        &$\times10^{-3}$&                 &                     &                        &                      & $1.5\pm0.1$              \\
$kT_{DBB}$      & $\flat$    & $1.62\pm0.01$   & $1.60\pm0.01$       & $0.89^{+0.04}_{-0.12}$ & $1.02\pm0.01$        & $1.34\pm0.05$            \\
$N_{DBB}$       & $\ddag$    & $5.3\pm0.1$     & $5.6\pm0.1$         & $6.6^{+0.8}_{-4.0}$    & $8.3\pm0.1$          & $8.7\pm0.5$              \\
$kT_{BB}$       & $\flat$    & $3.06\pm0.03$   & $2.96\pm0.03$       & $1.23\pm0.05$          & $1.55\pm0.01$        & $2.51^{+0.05}_{-0.11}$   \\
$N_{BB}$        & $\ddag$    & $0.30\pm0.01$   & $0.37\pm0.02$       & $5.2\pm1.5$            & $1.96\pm0.05$        & $0.43\pm0.1$             \\
$kT_e$          &            &                 &                     & $3.72\pm0.05$          & $3.77\pm0.02$        & $4.4^{+0.7}_{-0.5}$      \\
$N_{Comp}$   &$\times10^{-2}$&                 &                     & $7.9\pm1.5$            & $6.46\pm0.04$        & $2.61^{+1.1}_{-0.8}$     \\ 
$c_{MEG}$       &            & $1.03\pm0.01$   & $1.03\pm0.01$       & $1.03\pm0.01$          & $1.03\pm0.01$        & $1.03\pm0.01$            \\
$c_{FPMA}$      &            & $1.00\pm0.01$   & $1.00\pm0.01$       & $1.00\pm0.01$          & $1.00\pm0.01$        & $1.00\pm0.01$            \\
$c_{FPMB}$      &            & $1.00\pm0.01$   & $1.00\pm0.01$       & $1.00\pm0.01$          & $1.00\pm0.01$        & $1.00\pm0.01$            \\
$Cstat/dof$     &            & 9777.8/7094     & 8133.6/7092         & 7935.7/7090            & 7751.4/7090          & 7584.7/7088              \\
\hline
\hline
\end{tabular} 
\caption{Best fit parameters of the simultaneous \chandra\ HETG and \nustar\ spectra.
The \chandra\ HEG and MEG first order as well as \nustar\ FPMA and FPMB spectra 
(in black, red, green and blue, respectively) are fit simultaneously, 
leaving a cross-normalisation constant free to vary ($c_{MEG}$, $c_{nuA}$ and $c_{nuB}$). 
The first model, B1, {\sc DBB+BB} is composed by {\sc tbabs(diskbb+bbody)}, 
the second, B2, is additionally absorbed by an approximate ionised absorption 
component ({\sc tbabs*zxipcf(diskbb+bbody)}). Model B3 additionally considers 
the emission from a thermal Comptonisation component 
({\sc tbabs*zxipcf(diskbb+bbody+nthcomp)}). The continuum emission 
components in model B6 correspond to the ones of B3, however the ionised 
absorption is substituted by a self consistent one ({\sc $IA_{soft}$}: 
{\sc tbabs*IA(diskbb+bbody+nthcomp)}). Finally, model B7 additionally consider 
the emission from a relativistic Fe~K$\alpha$ line 
({\sc tbabs*IA(diskbb+bbody+nthcomp+diskline)}).
See text for description of the various parameters. $\flat$ In keV units. 
$\star$ In units of $10^{22}$ cm$^{-2}$. 
$\ddag$ Normalisation $N=R^2 cos(\theta)$, where $R^2$ is the apparent 
disc inner disc radius (or blackbody radius) in km, and $\theta$ the angle 
of the disc ($\theta=0$ face on, $cos(\theta)=1$ for blackbody). 
}
\label{Nutab}
\end{center}
\end{table*} 

The observed features (see Fig. \ref{NuHETGSPEC} and 
\S~\ref{HETGspec}) indicate the presence of ionised absorption. 
To properly compute the model of the ionised absorption component, 
the knowledge of the irradiating SED is required. 
We, therefore, started by first adding an "approximated" ionised absorption 
component ({\sc zxipcf}), with the main purpose to initially 
derive the fiducial SED to be used as input for the photo-ionisation computations. 
After the computation of the proper absorption model the spectra will be 
re-fitted, obtaining the final best fit. 

The addition of an ionised absorption component ({\sc B2: tbabs*zxipcf(diskbb+bbody)}) 
significantly improved the fit ($\Delta C-stat=1616.9$ for the addition of 2 parameters). 
The ionised absorption component reasonably reproduced both the soft X-ray 
absorption lines as well as the strong features in the Fe K band with a large 
column density ($log(N_H/cm^{-2})=23.6\pm0.1$) of highly ionised material 
($log(\xi/erg~cm~s^{-1})=4.25\pm0.02$; Tab. \ref{Nutab}). However, highly significant residuals 
were still present (Fig. \ref{NuHETGSPEC}). 

We noted that the excess of emission at energies above $E\geq25$~keV is 
likely the signature of an additional Comptonisation component. 
We therefore added to the model a thermal inverse-Comptonisation 
component ({\sc nthcomp} in {\sc Xspec}). We assumed an asymptotic 
photon index of $\Gamma=2$ and seed photons produced 
by the disc black-body. 
The fit with this model ({\sc B3: tbabs*zxipcf(diskbb+bbody+nthcomp)}) 
provided a significant improvement of the fit ($\Delta C-stat=197.8$ 
for the addition of 2 free parameters; Tab. \ref{Nutab}), properly 
reproducing the high energy emission (Fig. \ref{NuHETGSPEC}).
However, very significant residuals were still present in the $\sim7-10$~keV 
energy range. We noted that part of those residuals might be due 
to the inaccurate modelling of the ionised absorber. 
Indeed, the ionised Fe~K edges at $E_{Fe {\sc xxv}}=8.83$ and 
$E_{Fe {\sc xxvi}}=9.28$~keV can contribute to the observed residuals. 
We, therefore, defer the discussion of such residuals until after the computation 
of the proper ionised absorber model (see \S~\ref{Fitcha} and \ref{FitBB}). 

\subsection{Soft state SED}

\begin{figure}
\begin{center}
\vspace{-0.2cm}
\includegraphics[height=0.35\textwidth,angle=-0]{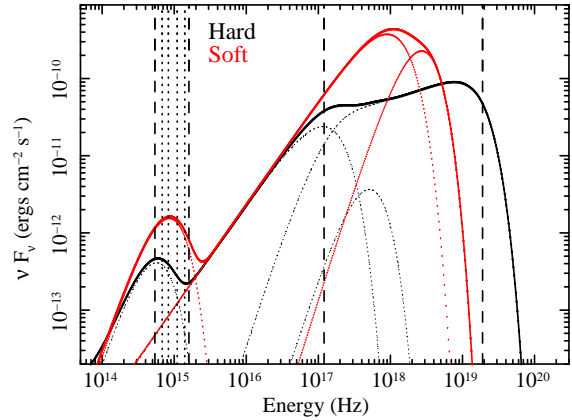}
\caption{In black and red are shown the de-absorbed best fit "bona fide" 
spectral energy distribution during the hard and soft state, respectively. 
The dashed lines show the extremes of the X-ray and optical-UV energy 
ranges constrained by \nustar, \xmm, \chandra\ and the \swift\ filters. 
The dotted vertical lines show the mean energy of the various filters. }
\label{SED}
\end{center}
\end{figure}
The red solid line in Fig. \ref{SED} shows the de-absorbed best fit (model 
B2 of Tab. \ref{Nutab}) soft state SED. The dotted red lines peaking at 
$\sim3\times10^{17}$ and $\sim1.5\times10^{18}$~Hz show the disk 
black-body and the black-body components, respectively. 
The dashed vertical lines indicate the boundaries where the SED is 
observationally constrained by the \chandra+\nustar\ data at high 
energy and by the \swift-UVOT observations (performed in various 
filters) in the optical-UV band. 
The observed optical-UV magnitudes range between 
$m_{V_{AB}}\sim18$ to $m_{UVW2_{AB}}\sim20$. This 
optical-UV flux is significantly higher than the extrapolated 
emission from the disc black body component dominating in 
the X-ray band (see Fig. \ref{SED}). Therefore, this extra flux indicates 
the presence of an additional component, possibly associated 
with the irradiated outer disc or the companion star (Hynes et al. 2002; 
Migliari et al. 2010). We reproduced this emission by adding 
a black-body component peaking at $\sim6\times10^{14}$~Hz 
(see Fig. \ref{SED}). 

We used such bona-fide SED as input for the photo-ionisation 
computations of the absorber that will be presented in \S~\ref{Fitcha}
and \ref{FitBB}. 
Additionally, we tested that considering any of the best fit continuum 
models in Tab. \ref{Nutab} produced negligible effects on the ionised 
absorber photo-ionisation stability and properties. 
The photo-ionisation computations have been 
performed with {\sc cloudy} 17.00 (Ferland et al. 2013). 

\section{Absorption lines: \chandra\ HETG spectrum}
\label{HETGspec}

\begin{figure*}
\vspace{-1.3cm}

\hspace{-1.8254cm}
\includegraphics[height=1.1\textwidth,width=0.9\textwidth,angle=90]{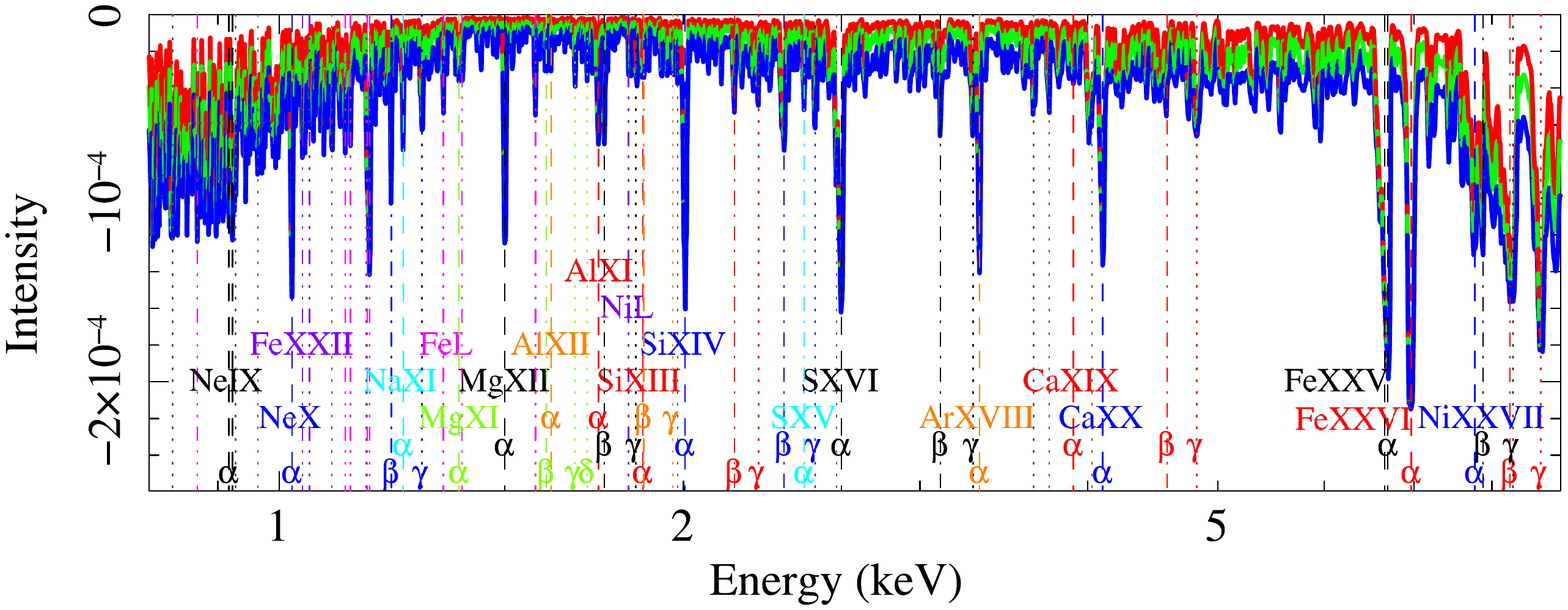}
\vspace{-7.7cm}
\caption{Residuals to the presence of narrow ($\sigma=1$~eV) 
absorption lines in the \chandra\ HETG (HEG+MEG) spectra. 
The red, green and blue solid lines show the 68, 90 and 99 \% 
confidence contours. Sixty absorption lines are detected. 
We highlight with a vertical line each detected absorption feature. 
The same colour is used for features associated to the same series. 
In particular, the energies of the $\alpha$, $\beta$ and $\gamma$ 
transitions are reported with dashed, dash-dotted and dotted lines, 
respectively. Absorption lines associated with Fe~L features are 
highlighted with magenta dot-dot-dot-dashed lines (in violet is 
the density sensitive Fe~{\sc xxii} doublet). Unidentified lines are 
shown with grey dotted lines. }
\label{Cha}
\end{figure*}
We started the characterisation of absorption during the soft state 
observation, by performing a blind search to hunt for narrow 
absorption lines in the spectrum of the persistent emission 
(see \S 9.4 for a discussion of possible caveats). 
We fitted the \chandra\ grating data (HEG 
and MEG are fitted simultaneously) with the 
best fit continuum model reproducing the \chandra+\nustar\ 
continuum and we added a narrow Gaussian absorption line.
We then computed the confidence contours 
by stepping the energy of the line from $E=0.8$ to $9$~keV 
(with 1640 steps logarithmically equally spaced) and the intensity 
in 10 steps from 0 to $-3\times10^{-4}$~photons~cm$^{-2}$~s$^{-1}$ 
in the line. The red, green and blue contours in Fig. \ref{Cha}
show the 68, 90 and 99 \% confidence levels. 
This plot clearly shows the detection of many very significant 
absorption lines (Fig. \ref{Cha}). 

To more accurately determine the significance of the detection 
of the expected absorption features and to measure the equivalent 
width of the lines, we first divided the spectrum into 
intervals of $\sim2$~\AA~width and fitted it with the same continuum 
model (allowing for variations of the normalisation). We then added to 
the model a narrow absorption line with Gaussian profile for each 
observed residual. 
We then tested the significance of the residual and we reported 
in Tab. \ref{HETGlines} all lines detected at more than 
90~\% significance, with associated velocity shifts and 
equivalent widths. We also highlight with bold characters 
in Tab. \ref{HETGlines} the lines detected at more than $3~\sigma$.

Sixty lines are significantly detected, of which 
we identified 51. Of these 21 lines are detected at more than $3~\sigma$.
The strongest lines are due to the Ly-like and He-like transitions 
of many elements such as: Ne, Na, Mg, Al, Si, S, Ar, Ca, Fe, and Ni, 
as well as lines of the Fe~L and Ni~L complexes. 
Interestingly, for several elements the entire series from 
the $\alpha$ to the $\gamma$ lines are detected and identified. 
Wherever possible we separately reported the equivalent widths 
of the inter-combination and resonance lines of the He$\alpha$ 
triplets. 

Below $\sim1.8$~keV, we detected several absorption features 
consistent with Fe L transitions including the Fe XXII density 
sensitive doublet. In particular, the lower energy line of the 
Fe XXII doublet ($\lambda=11.920$) is formally detected at 
more than $2.5$~$\sigma$ confidence, while the higher energy transition 
(at $\lambda=11.77$) is detected with a significance just above 
$1~\sigma$. Therefore, we fixed its energy to the expected 
value\footnote{We also tried fixing the energy of this transition 
to that expected if the line experienced the same shift as the 
$\lambda=11.920$ line (\S~6.2), which did not change our result. }  
and we reported the $1~\sigma$ uncertainties in Tab. \ref{HETGlines}. 

We marked with a question mark the Al XII He$\beta$ and $\gamma$ 
lines, because both have intensities close to the detection limit. 
Additionally, the Al XII He$\beta$ line is affected by the wings 
of the more intense Si XIII He$\alpha$ line. For these reasons, 
we could not robustly constrain its energy through fitting, and therefore 
chose to fix its energy to the theoretical value. We also marked with a question 
mark the $\alpha$, $\beta$, $\gamma$ and $\delta$ lines 
of the Mg XI He-like series. Indeed, we observed that the higher 
order transitions (e.g, $\gamma$ and $\delta$) are detected at 
low significance, but have strengths 
and equivalent widths comparable to or higher than the respective 
$\alpha$ line. We considered this doubtful. 
The same occurred for the Ca XIX He$\alpha$, $\beta$ and 
$\gamma$ lines. 

We note that nine lines remained unidentified. A few of these 
might be associated with spurious detections. 

\begin{table}
{\scriptsize
\begin{tabular}{ l c c c c }
\hline
\hline
Identification & $\lambda$   & $E_O$ & $v_{out}$   & EW \\
               & (\AA)       & (keV) & (km s$^{-1}$) & (eV) \\
\hline
Fe XVIII L             & 14.28 & $0.869\dag$                   &             & $-0.66\pm0.43$ \\
Ne IX He$\alpha$ (i)   & 13.553& $0.917\pm0.005$           &             & $-0.58\pm0.47$ \\
Ne IX He$\alpha$ (r)   & 13.445& $0.923\pm0.007$           &             & $-0.56\pm0.38$ \\
Unid                   &       & $0.928\pm0.006$           &             & $-0.57\pm0.39$ \\
Unid                   &       & $0.964^{+0.007}_{-0.017}$ &             & $-0.37\pm0.31$ \\
{\bf Ne X Ly$\alpha$}        & 12.133& $1.0222\pm0.0004$         & $-90\pm120$ & $-1.72\pm0.25$ \\
{\color{blue}Fe XXII L $1~\sigma$}&11.92& $1.0405^{+0.002}_{-0.009}$&  & $-0.38^{+0.15}_{-0.27}$ \\
{\color{blue}Fe XXII L $1~\sigma$}&11.77 & 1.0537           &           & $-0.28^{+0.19}_{-0.23}$ \\
Unid                   &       & $1.094^{+0.007}_{-0.007}$ &             & $-0.36\pm0.31$ \\
Fe XVII                & 11.018& $1.120\pm0.007$           &             & $-0.57\pm0.38$ \\
Fe XXIII               & 10.981& $1.130\pm0.005$           &             & $-0.49\pm0.34$ \\
Unid                   &       & $1.161\pm0.003$           &             & $-0.50\pm0.39$ \\
{\bf Fe XXIV L}              & 10.663& $1.1632\pm0.0005$         & $-100\pm130$& $-0.97\pm0.31$ \\
{\bf Fe XXIV L}              & 10.619& $1.1678\pm0.00026$        & $-50\pm70$  & $-1.53\pm0.24$ \\
{\bf Ne X Ly$\beta$}         & 10.239& $1.211\pm0.004$           &             & $-0.66\pm0.27$\\
{\bf Na XI Ly$\alpha$}       & 10.025& $1.237^{+0.015}_{-0.007}$ &             & $-0.39\pm0.22$\\
Ne X Ly$\gamma$        & 9.7082& $1.277^{+0.004}_{-0.015}$ &             & $-0.26\pm0.23$\\
Fe XXI L               & 9.356 & $1.325^{+0.006}_{-0.09}$  &             & $-0.30\pm0.21$ \\
Mg XI He$\alpha$?      & 9.1688& $1.361^{+0.007}_{-0.010}$ &             & $-0.33\pm0.24$ \\
Fe XXII L              & 9.057 & $1.368^{+0.004}_{-0.009}$ &             & $-0.25\pm0.18$ \\
{\bf Mg XII Ly$\alpha$}      & 8.4210& $1.4728\pm0.0003$         & $-100\pm60$ & $-2.91\pm0.21$ \\
{\bf Fe XXIV L}              & 7.9893& $1.552\pm0.002$           &             & $-0.38\pm0.18$ \\
Mg XI He$\beta$?       & 7.852 & $1.581\pm0.006$           &             & $-0.29\pm0.23$ \\
Al XII He$\alpha$      & 7.778 & $1.594^{+0.010}_{-0.018}$ &             & $-0.25\pm0.20$ \\
Mg XI He$\gamma$?      & 7.473 & $1.660\pm0.001$           & $-180\pm180$& $-0.38\pm0.23$ \\
Mg XI He$\delta$?      & 7.310 & $1.695\pm0.002$           & $170\pm350$ & $-0.39\pm0.23$ \\
{\bf Al XIII Ly$\alpha$}     & 7.1727& $1.7284\pm0.0009\ddag$    & $40\pm160$  & $-1.25\pm0.28$ \\
{\bf Mg XII Ly$\beta$}       & 7.1062& $1.7458\pm0.0005$         & $-190\pm90$ & $-0.80\pm0.20$ \\
Ni XXVI L              & 6.8163& $1.820^{+0.012}_{-0.001}$ &             & $-0.24\pm0.22$ \\
{\bf Mg XII Ly$\gamma$}      & 6.7379& $1.843\pm0.007$           &             & $-0.49\pm0.26$ \\
Si XIII He$\alpha$     & 6.6480& $1.865\pm0.009$           &             & $-0.34\pm0.26$ \\
Al XII He$\beta$?      & 6.6348& $1.8687\dag$              &             & $-0.26\pm0.25$ \\
Al XII He$\gamma$?     & 6.3129& $1.964\dag$               &             & $-0.21\pm0.19$ \\
Unid                   &       & $1.980\pm0.004$           &             & $-0.42\pm0.27$ \\
{\bf Si XIV Ly$\alpha$}      & 6.1822& $2.0058\pm0.0003$         & $-40\pm45$  & $-5.58\pm0.31$ \\
Si XIII He$\beta$      & 5.680 & $2.183\dag$                   &             & $-0.43\pm0.41$ \\
Si XIII He$\gamma$     & 5.405 & $2.275\pm0.015$           &             & $-0.71\pm0.42$ \\
{\bf Si XIV Ly$\beta$}       & 5.2172& $2.375^{+0.004}_{-0.003}$ & $190\pm500$ & $-1.11\pm0.60$ \\
S XV He$\alpha$        & 5.0387& $2.460\pm0.003$           & $70\pm360$  & $-0.83\pm0.52$ \\
Si XIV Ly$\gamma$      & 4.9469& $2.51\pm0.01$             &             & $-0.66\pm0.63$ \\
{\bf Unid}                   &       & $2.597^{+0.004}_{-0.010}$ &             & $-1.43\pm0.78$ \\
{\bf S XVI Ly$\alpha$}       & 4.7292& $2.621\pm0.001$           & $110\pm110$ & $-6.49\pm0.72$ \\
{\bf S XVI Ly$\beta$}        & 3.9912& $3.105^{+0.010}_{-0.004}$ &             & $-1.38\pm0.77$ \\
S XVI Ly$\gamma$       & 3.7845& $3.286^{+0.008}_{-0.017}$ &             & $-1.37\pm0.85$ \\
{\bf Ar XVIII Ly$\alpha$}    & 3.7329& $3.323\pm0.002$           & $-180\pm180$& $-4.38\pm0.94$ \\
Unid                   &       & $3.645^{+0.021}_{-0.004}$ &             & $-1.2\pm0.9$  \\
Unid                   &       & $3.744^{+0.018}_{-0.009}$ &             & $-1.3\pm1.1$  \\
Ca XIX He$\alpha$?     & 3.1772& $3.90\pm0.01$             &             & $-0.9\pm0.8$  \\
{\bf Unid}                   &       & $4.029\pm0.008$           &             & $-2.16\pm1.00$ \\
{\bf Ca XX Ly$\alpha$}       & 3.0203& $4.106\pm0.004$           & $-70\pm300$ & $-7.16\pm1.55$ \\
Ca XIX He$\beta$?      & 2.7054& $4.583\pm0.03$            &             & $-1.83\pm1.76$ \\
Ca XIX He$\gamma$?     & 2.571 & $4.821\pm0.02$            &             & $-2.31\pm1.86$ \\
{\bf Fe XXV He$\alpha$ (i)}  & 1.857 & $6.656\pm0.008$           &             & $-15.1\pm4.4$ \\
{\bf Fe XXV He$\alpha$ (r)}  & 1.8504& $6.692\pm0.007$           & $370\pm300$ & $-33.0\pm3.8$ \\
{\bf Fe XXVI Ly$\alpha$}     & 1.7798& $6.961\pm0.004$           & $220\pm170$ & $-45.8\pm3.5$ \\
Ni XXVII He$\alpha$    & 1.596 & $7.78\pm0.04$             &             & $-15\pm10$ \\
FeXXV He$\beta$        & 1.5732& $7.86\pm0.07$             &             & $-13\pm10$   \\
Fe XXVI Ly$\beta$      & 1.5028& $8.249\dag$                   &             & $-14\pm11$   \\
FeXXV He$\gamma$       & 1.495 & $8.293\dag$                   &             & $-16\pm10$   \\
Fe XXVI Ly$\gamma$     & 1.425 & $8.709\pm0.015$           &             & $-48.5\pm35$ \\
\hline
\hline
\end{tabular} 
\caption{Best fit energies ($E_O$), equivalent widths ($EW$) and outflow 
velocity ($v_{out}$) of lines detected in the HETG (HEG+MEG) spectra.
For each identified transition the expected wavelength of the transition 
($\lambda$) is reported. The energies of the doublets are averaged 
over the oscillator strengths. We only list velocities for the strongest 
transitions. $\dag$ The energy of these transitions 
have been fixed, in the fit, to the expected values. }
\label{HETGlines}
}
\end{table} 

\begin{figure*}
\vspace{-0.5cm}
\includegraphics[height=0.65\textwidth,angle=0]{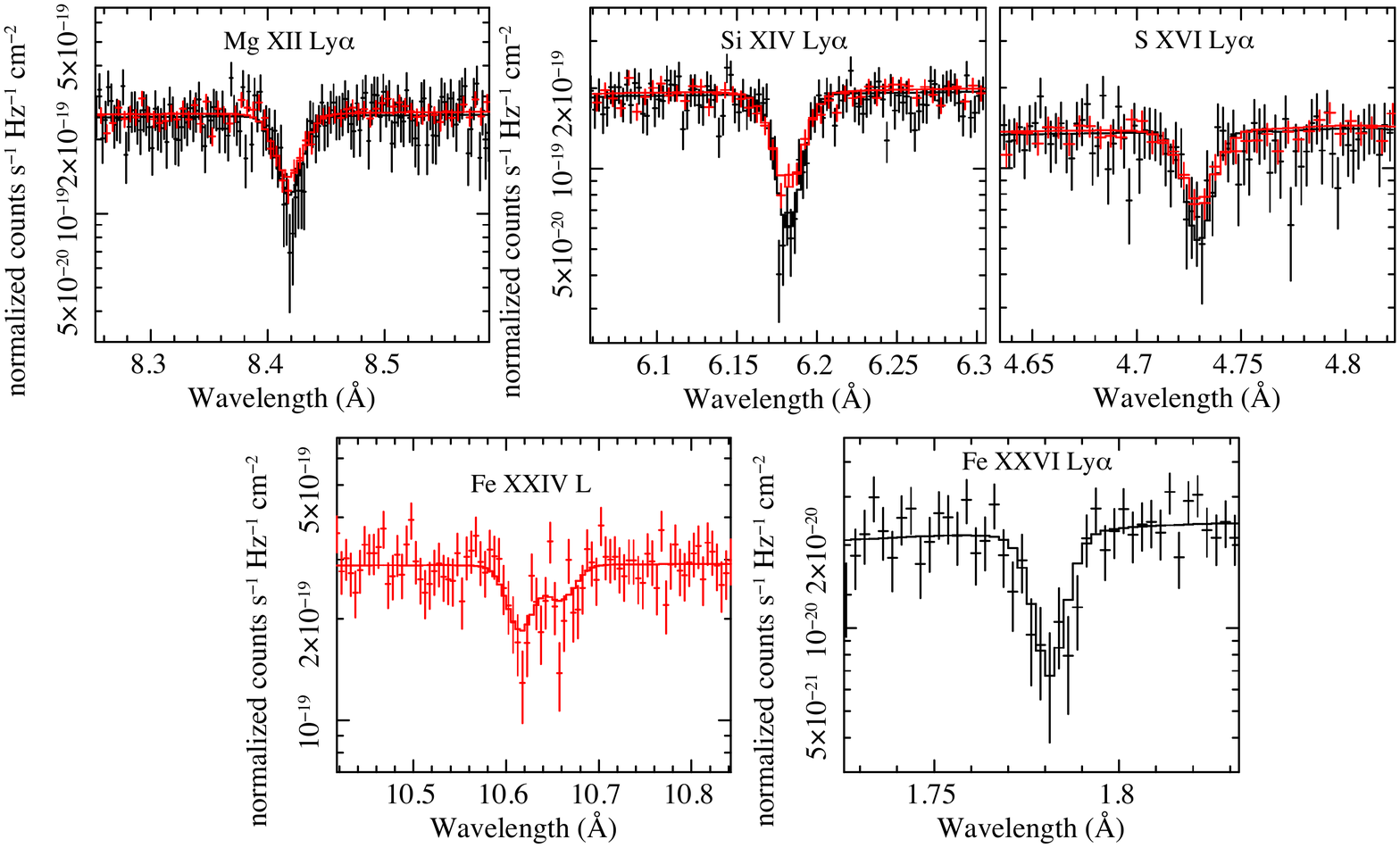}
\vspace{-1.3cm}
\caption{The red and black data display the \chandra\ HETG MEG and HEG 
spectra at the energies of some of the strongest lines, respectively, 
fitted with the best fit model (B7 in Tab. 1). {\it (Top panels)} from left 
to right show the Mg {\sc xii} Ly$\alpha$, Si {\sc xiv} Ly$\alpha$ and 
S {\sc xvi} Ly$\alpha$ lines. The profile of all lines is well reproduced 
by the best fit photo-ionised absorption component. 
{\it (Bottom panels)} from left to right show the two lines of 
the Fe {\sc xxiv} L complex and the Fe {\sc xxvi} Ly$\alpha$ 
transition. Only data of the MEG and HEG instruments are shown 
because of their higher effective area at low and high energies, 
respectively. }
\label{FitLines}
\end{figure*}

\subsection{The density sensitive \Fetwt\ doublet}
\label{doublet}

\begin{figure}
\begin{center}
\vspace{-0.3cm}
\includegraphics[height=0.45\textwidth,angle=0]{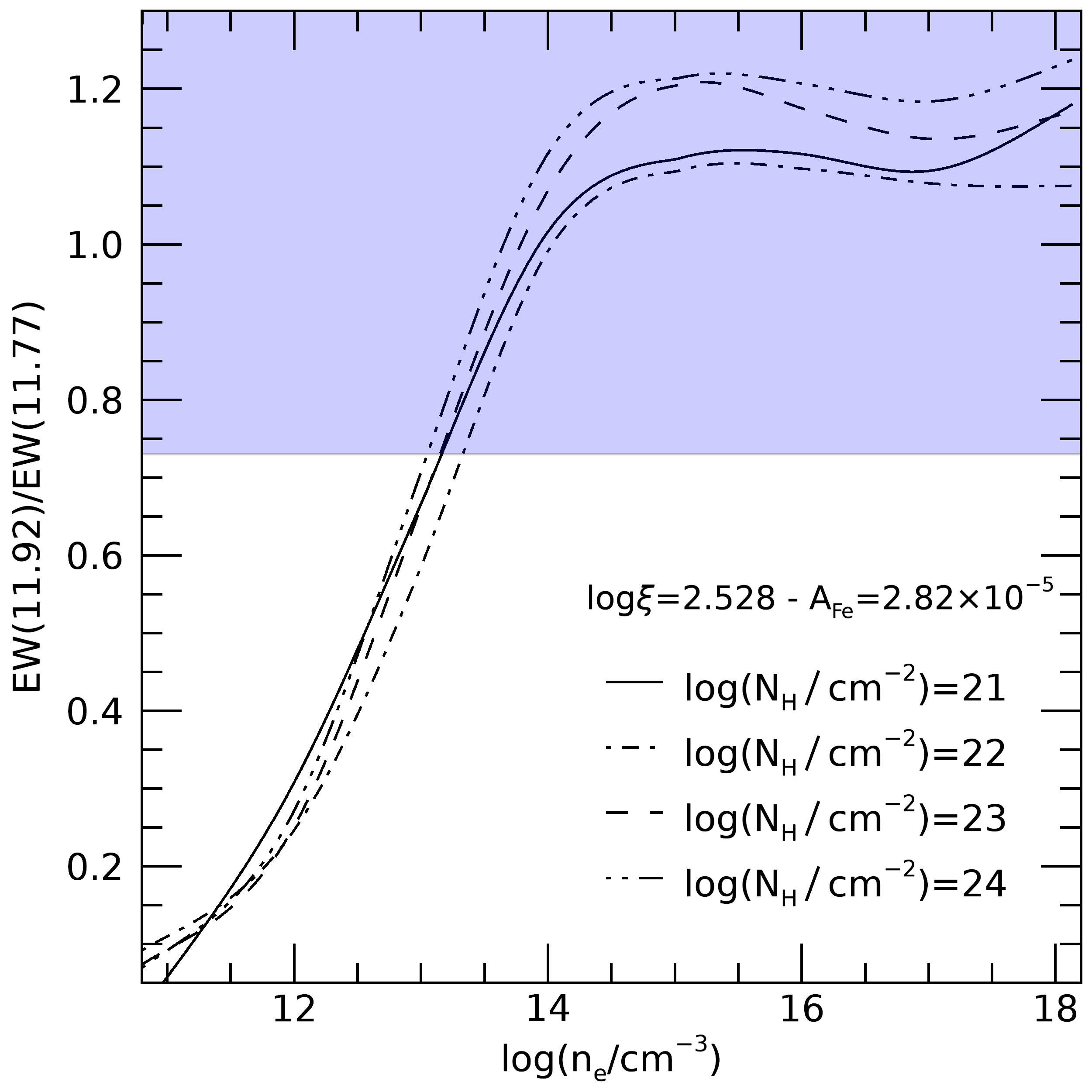}
\caption{Ratio of equivalent widths of the absorption lines 
of the Fe~XXII~L doublet ($EW(11.92)/EW(11.77$\AA$)$) 
as a function of the plasma density ($n$). The various lines 
indicate the relation between plasma density and line ratio 
for different column densities of the absorber. The violet area 
indicates the region consistent (at $1~\sigma$) with the observed 
line ratio. }
\label{Fexxii}
\end{center}
\end{figure}
The ratio of the intensities of the Fe~XXII~L lines at $\lambda$ 
11.92~\AA\ and 11.77~\AA\ is a good density diagnostic 
(Mauche et al. 2003). The \Fetwt\ line at 11.92 \AA\ 
was observed with an equivalent width of $-0.38^{+0.15}_{-0.27}$~eV 
($1~\sigma$), while the line at 11.77 \AA\ was barely detected 
with EW$=-0.28^{+0.19}_{-0.23}$~eV. The uncertainty 
on the line ratio was estimated by assuming that both lines have 
the same widths and that their equivalent widths follow the relation: 
$EW_{11.92} = f \times EW_{11.77}$. The violet region in Fig. 
\ref{Fexxii} represents the $1~\sigma$ uncertainty on the line ratio
(computing the 1~$\sigma$ error on the value of $f$ 
directly from the spectrum), corresponding with 
$EW_{11.92} > 0.73 EW_{11.77}$. 

By performing extensive {\sc cloudy} simulations, we computed 
the doublet line ratio as a function of the plasma density for various 
values of the column density of the absorbing material 
(see solid, dashed and dot-dashed lines in Fig. \ref{Fexxii}). 
We confirmed that the ratio is a sensitive probe 
of the plasma density. 
In particular, the observed line ratio suggests 
a plasma density of $n>10^{13}$~cm$^{-3}$. This result will be 
used in \S~\ref{distance} to estimate the location of the ionised absorber. 

\subsection{Line shifts}

The absorption line centroids in \mxb\ are observed to shift with orbital 
phase, with a semi-amplitude of $\sim90$~km~s$^{-1}$ (Ponti et al. 2018). 
Such modulation is thought to trace the radial velocity curve of the primary. 
We note that the HETG observation analysed here covered two complete 
orbital periods. Therefore, we expect to see no systematic shift of the 
absorption lines, due to such orbital modulation (the main consequence 
being an artificial broadening of the lines). 
We also note that Sidoli et al. (2001) investigated the evolution of 
the \Fevc\ and \Fevs\ line intensities as a function of orbital phase, 
finding no significant variation (despite intensity variations as large as 
a factor of two were not excluded). 

The line centroids of all lines in the average spectrum are consistent 
with being at rest, with outflow or inflow velocities lower than 
$\sim200$~km~s$^{-1}$ (Tab. \ref{HETGlines}). 
In particular, we observed that the lines with the highest signal to noise 
indicated upper limits to any bulk flow velocity of less than 
$\sim50-70$~km~s$^{-1}$ (Tab. \ref{HETGlines}). Therefore, the observed 
ionised absorption is not due to a wind, it is instead associated 
with an ionised disc atmosphere.

\subsection{The line-rich HETG spectrum fitted with self-consistent photo-ionisation models}
\label{Fitcha}

In this paragraph, we present the separate fit of only the \chandra\ HETG 
spectrum (0.8-7.3~keV; Tab. 3). 
In this way, the fit will be driven by the absorption lines only (e.g., without 
strong contamination from either the ionised Fe~K edges, the broad Fe~K line 
or the shape of the high energy continuum). 

As already evidenced from \S~\ref{BBfit}, we observed that the addition 
of an approximated ionised absorption component (model {\sc C2}) 
to the disk black-body and hot black-body emission ({\sc C1}) significantly 
improved the fit of the HETG spectrum of \mxb, during the soft state 
($\Delta C-stat=1549.2$ for the addition of 2 new free parameters; 
see Tab. \ref{Chatab}). 
Based on the observed soft state SED, we built a fully auto consistent 
photo-ionisation model (IA$_{\rm soft}$). The model table was computed 
through a {\sc cloudy} computation, assuming constant 
electron density ($n_e=10^{14}$~cm$^{-3}$), turbulent velocity 
$v_{turb}=500$~km~s$^{-1}$ and Solar abundances. 

We then substituted the approximated photo-ionised component with this 
self-consistent ionised absorber ({\sc C4}). By comparing model {\sc C4} 
with {\sc C2}, we observed a significant improvement of the fit ($\Delta C-stat=207.3$) 
for the same degrees of freedom (Tab. \ref{Chatab}). 
This confirms that the array of absorption lines is better described 
by the self-consistent photo-ionisation model, producing an acceptable 
description of the data. The ionised plasma is best described by 
a relatively large column density ($log(N_H/cm^{-2})=23.6\pm0.1$) 
of highly ionised ($log(\xi/erg~cm~s^{-1})=3.85^{+0.04}_{-0.11}$) material. 
We note that these values are in line with what measured 
from \xmm\ observations of \mxb\ during the previous outburst 
(Sidoli et al. 2001; Diaz Trigo et al. 2006). 

We then tested whether the ionised absorber is composed of multiple 
and separate components 
with, for example, different ionisation parameters. 
We performed this by adding a second ionised absorber 
layer to the fit. We observed a slight improvement of the fit 
($\Delta C-stat=29.0$ for the addition of 2 free parameters; 
Tab. \ref{Chatab}). The two components of the ionised absorbers 
were split into a very high column density and ionisation parameter 
component ($log(N_H/cm^{-2})\sim24.2$ and 
$log(\xi/erg~cm~s^{-1})>4.5$) and a much lower 
column density and ionisation one ($log(N_H/cm^{-2})<22.5$ and 
$log(\xi/erg~cm~s^{-1})\sim3.3$), leaving the best fit parameters of the continuum 
emission consistent with the previous fit (Tab. \ref{Chatab}). 

Since the fit with two ionised absorption layers provided only 
a marginal improvement of the fit from a statistical point of view, 
we concluded that a single ionised absorption layer provided 
a good description of the absorption lines. 

\begin{table*}
\begin{center}
\begin{tabular}{ l l c c c c c c c c c c c c c c c c}
\hline
\hline
\multicolumn{6}{c}{\bf Only \chandra\ HETG soft state (2016-04-21)} \\
\hline
\hline
Model           &            & {\sc C1}       & {\sc C2}               & {\sc C4}               & {\sc C5}             \\
                &            & {\sc DBB+BB}   & {\sc xipcf(DBB+BB)}    & {\sc IA(DBB+BB)}       & {\sc IA*IA(DBB+BB)}  \\
\hline				
$N_{H}$         &$\star$     & $0.23\pm0.01$  & $0.22\pm0.02$          & $0.213\pm0.008$        & $0.192\pm0.007$      \\
$log(\xi_{1})$  &            &                & $4.26\pm0.02$          & $3.85^{+0.04}_{-0.11}$ & $>4.56$              \\
$log(N_{H_{1}})$&$\star$     &                & $23.6\pm0.1$           & $23.6\pm0.1$           & $24.21\pm0.02$       \\
$log(\xi_{2})$  &            &                &                        &                        & $3.33\pm0.04$        \\
$log(N_{H_{2}})$&$\star$     &                &                        &                        & $<22.5$              \\
$kT_{DBB}$      & $\flat$    & $1.07\pm0.10$  & $1.2\pm0.2$            & $1.34\pm0.07$          & $1.34\pm0.07$        \\
$N_{DBB}$       & $\ddag$    & $19\pm6$       & $12\pm4$               & $9.9^{+2.4}_{-0.9}$    & $9.9^{+1.8}_{-1.1}$  \\
$kT_{BB}$       & $\flat$    & $1.7\pm0.2$    & $2.1^{+1.0}_{-0.3}$    & $2.8^{+0.1}_{-0.5}$    & $2.8\pm0.5$          \\
$N_{BB}$        & $\ddag$    & $5.0\pm2.0$    & $2.1\pm1.5$            & $0.8^{+0.1}_{-0.4}$    & $0.83\pm0.5$         \\
$c_{MEG}$       &            & $1.03\pm0.01$  & $1.03\pm0.01$          & $1.03\pm0.01$          & $1.03\pm0.01$        \\
$Cstat/dof$     &            & 8299.5/6137    & 6750.3/6129            & 6543.0/6129            & 6514.2/6127          \\
\hline
\hline
\end{tabular} 
\caption{Best fit parameters of the fit of the \chandra\ HETG spectra only.
See text for description of the various parameters. 
Symbols as in Tab. \ref{Nutab}. }
\label{Chatab}
\end{center}
\end{table*} 

\subsection{Stability curve}
\label{stability}

\begin{figure}
\includegraphics[height=0.48\textwidth,angle=0]{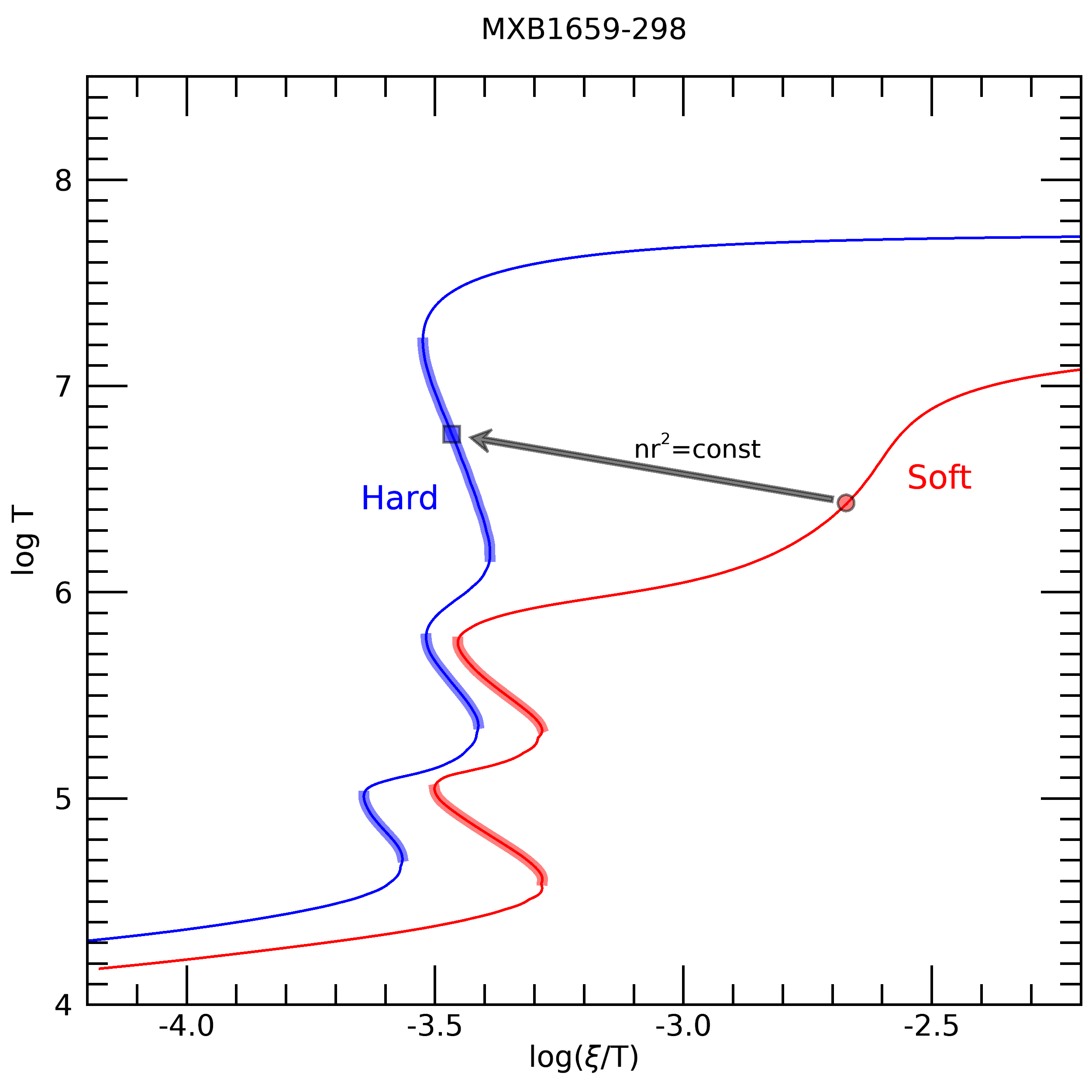}
\caption{Photoionisation stability curves of the ionised absorbing 
plasma when illuminated with the soft (red) and hard (blue) SEDs. 
The thick segments indicate the thermally unstable branches of the curves. 
The red circle indicates the position of the ionised absorber during the 
soft state. The absorber is thermally stable. The blue square indicates  
the expected condition of the absorber when illuminated by the hard 
state SED. Such conditions are thermally unstable. }
\label{Stability}
\end{figure}
The red line in Fig. \ref{Stability} shows the thermal equilibrium conditions 
of the absorber, once the ionised plasma is illuminated by the soft state 
SED (Fig. \ref{SED}). The thermal equilibrium is the result of the competition 
between heating and cooling and the corresponding stability curve is inferred 
through extensive {\sc cloudy} computations 
in the same way as described in Bianchi et al. (2017). 

As expected, we observed that the ionised absorber (i.e., disc atmosphere) 
is in thermal equilibrium during the soft state (see red point in Fig. \ref{SED}). 
Actually, any equilibrium solution with plasma temperatures higher 
than $kT\sim55$~eV or lower than $3.4$~eV would be stable, with only 
two small instabilities present between these two ranges. 

\section{Broad band fit of the soft state spectra} 
\label{FitBB}

We started the broad band fit of the soft state spectra of \mxb, by 
substituting the approximated photo-ionised absorber model ({\sc zxipcf})
employed in model {\sc B3} (\S \ref{BBfit}), with the fully auto consistent 
photo-ionised absorber model ({\sc B6}; {\sc 
tbabs*$IA_{soft}$(diskbb+bbody+nthcomp)}) described in \S  \ref{Fitcha}. 
We observed a significant improvement of the fit 
($\Delta C-stat=-184.3$ for the same dof), in agreement with the fact 
that the {\sc $IA_{\rm soft}$} model provides a superior description of the 
\chandra\ HETG spectra and the absorption lines. 
However, we noted that the best fit column density 
($log(N_H/cm^{-2})=24.25\pm0.02$ instead of $23.6\pm0.2$) 
and ionisation parameter ($log(\xi/erg~cm~s^{-1})=4.54\pm0.01$ 
instead of $3.85^{+0.04}_{-011}$) of the ionised 
absorber were significantly larger than the best fit of the HETG data only. 
This variation was induced by the attempt of the model to reproduce the large 
residuals at $\sim6-8$~keV (see Fig. \ref{NuHETGSPEC}). 
The resulting best fit was therefore driven to an increased column 
density and ionisation parameter of the absorbing plasma (compared 
with what was required by the absorption lines) in an attempt to enhance 
the depth of the ionised Fe~K edges ($E_{Fe {\sc xxv}}=8.83$ and 
$E_{Fe {\sc xxvi}}=9.28$~keV). 

We noted that the remaining significant residuals in the $\sim6-12$~keV 
band were resembling a broadened Fe~K$\alpha$ emission line 
(Fig. \ref{NuHETGSPEC}). We, therefore, added to the model a disk line 
component ({\sc B7; tbabs*$IA_{soft}$(diskline+diskbb+bbody+nthcomp)}). 
We assumed the line energy to be $E=6.4$~keV (although the material 
in the inner accretion disc might be ionised), a disk inclination 
of $75^\circ$ (Ponti et al. 2018), inner and outer radii of 
$r_{in}=6$~$r_{g}$ and $r_{out}=1000$~$r_{g}$ (where $r_g=GM_{NS}/c^2$ 
is the gravitational radius, $G$ is the Gravitational constant, $M_{NS}$ 
the NS mass and $c$ is the light speed).
The free parameters of the model were the disc emissivity index $\alpha_d$
and the line normalisation. This provided a significant improvement, 
resulting in an acceptable fit ($C-stat=7584.7$ for 7088 dof). 
The best fit line emissivity index and equivalent 
widths $\alpha_d\sim-2.6$ and $EW\sim250$~eV are consistent 
with the expected values from a standard irradiated accretion disc 
(Matt et al. 1991; George \& Fabian 1991). 

The addition of the disk-line component to the model allowed a better 
fit of the Fe~K band, decreasing the depth of the ionised Fe~K edges, 
therefore leading to best fit ionisation parameter and column density 
of the ionised absorber consistent with the fit of the HETG data alone
($log(N_H/cm^{-2})\sim23.5$~cm$^{-2}$ and $log(\xi/erg~cm~s^{-1})\sim3.8$). 

The observed and un-obscured 0.1-100 keV flux are 
$F_{0.1-100}=9.8\times10^{-10}$ and $11.2\times10^{-10}$
erg~cm$^{-2}$~s$^{-1}$, respectively. 
The disc black body, black body and Comptonisation components 
carry $\sim50$~\%, $\sim17$~\% and $\sim30$~\% of the 
un-obscured 0.1-100 keV flux. 
The best fit continuum is described by a disk black body 
component with temperature of $kT_{DBB}\sim1.3$~keV with an associated 
inner radius of $r_{DBB}\sim10$~km, therefore comparable with 
the NS radius. The best fit black body emission $kT_{BB}\sim2.5$~keV 
is hotter than the one of the accretion disc and is produced 
from a small region with a surface area of $r_{BB}\sim1.5$~km$^2$, likely 
associated with the boundary layer. 
At high energy ($E\geq20-30$~keV) a small contribution due to 
thermal Comptonisation appears significant. Assuming an asymptotic 
photon index of $\Gamma=2$, the best fit temperature of 
the Comptonising electrons is $kT_e\sim4$~keV. This value 
is significantly lower than what is typically observed during the hard state 
or in BH systems. This is possibly induced by the cooling power 
of the abundant soft photons produced by the disc and/or boundary 
layer (Done \& Gierlinski 2003; Burke et al. 2018).

\section{Hard state observations} 

To investigate the evolution of the disc atmosphere as a function of 
the accretion disc state and its response to the variation of the SED, 
we analysed the \xmm\ 
observation that caught \mxb\ in the hard state. 
The \nustar\ hard state observation (taken two days after) will be 
presented in a Degenaar et al. (in prep).

\label{xmm}

\begin{figure}
\hspace{-0.7cm}
\includegraphics[height=0.4\textwidth,angle=0]{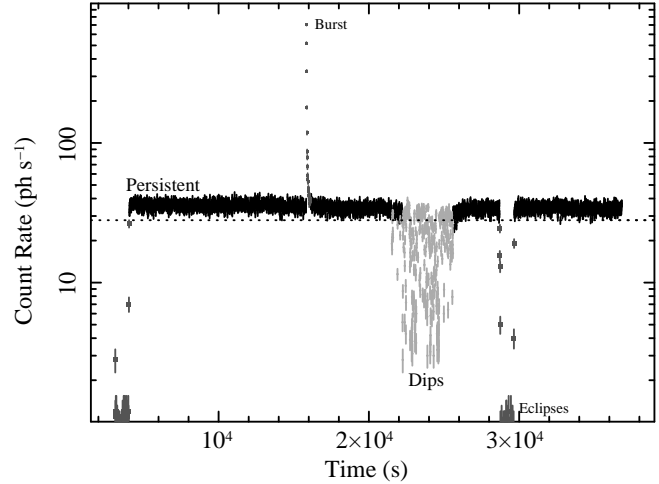}
\caption{0.5-10~keV \xmm\ light curve of \mxb\ performed on 2015-09-26. 
The EPIC-pn exposure starts just before the eclipsed period (shown with dark grey 
squares). During the observation a clear burst is observed (shown with dark grey 
colours) as well as intense dipping activity (light grey colours). The black points 
show the persistent emission. The dotted line indicates the lower bound we use 
to identify persistent flux, with intervals dipping below this being selected as dips. 
Time bis of 10~s are used. }
\label{xmmLC}
\end{figure}

Figure \ref{xmmLC} shows the light curve of the hard state \xmm\ 
EPIC-pn observation. 
One bright burst (dark grey peak), two eclipses (lasting $\sim930$~s) 
as well as intense dipping activity are observed and removed. 
\begin{figure}
\hspace{-0.05cm}
\includegraphics[height=0.4\textwidth,angle=0]{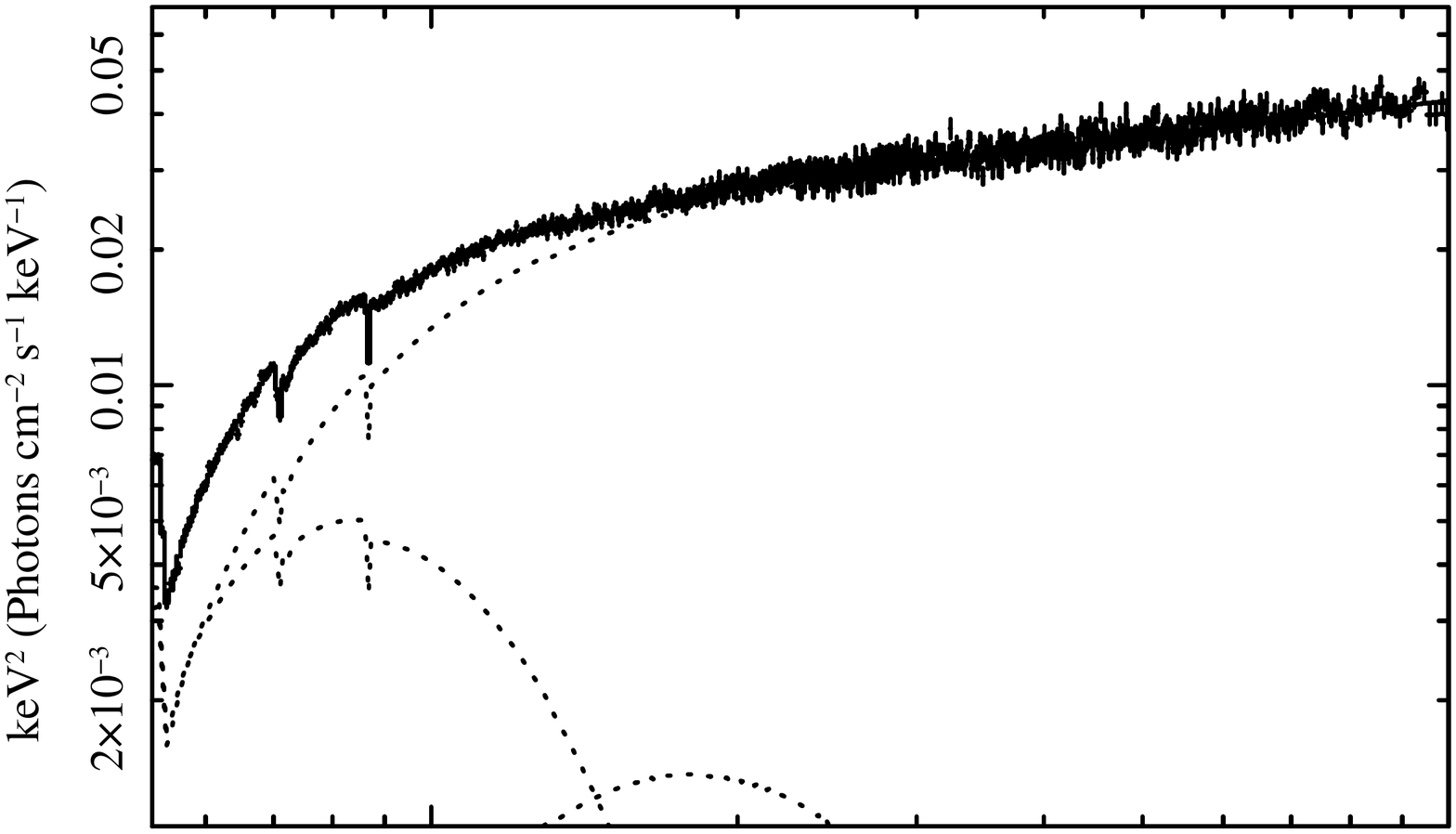}

\vspace{-5.72cm}
\includegraphics[height=0.4\textwidth,angle=0]{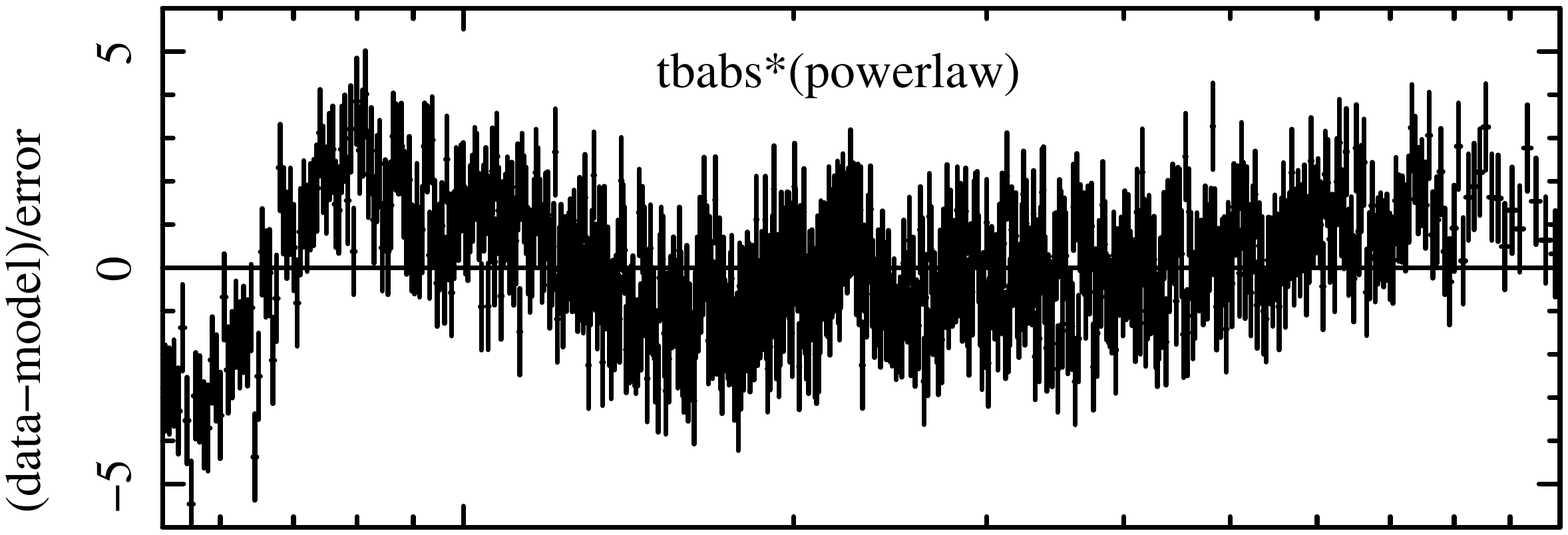}

\vspace{-4.52cm}
\includegraphics[height=0.4\textwidth,angle=0]{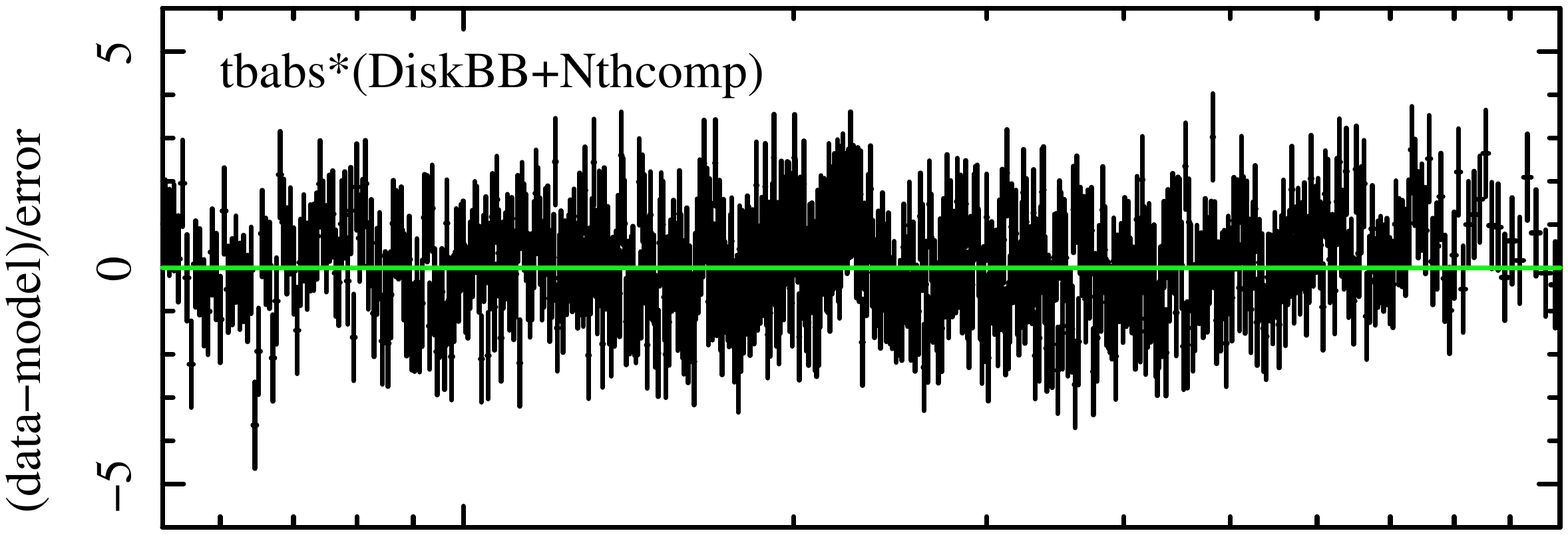}

\vspace{-4.52cm}
\includegraphics[height=0.4\textwidth,angle=0]{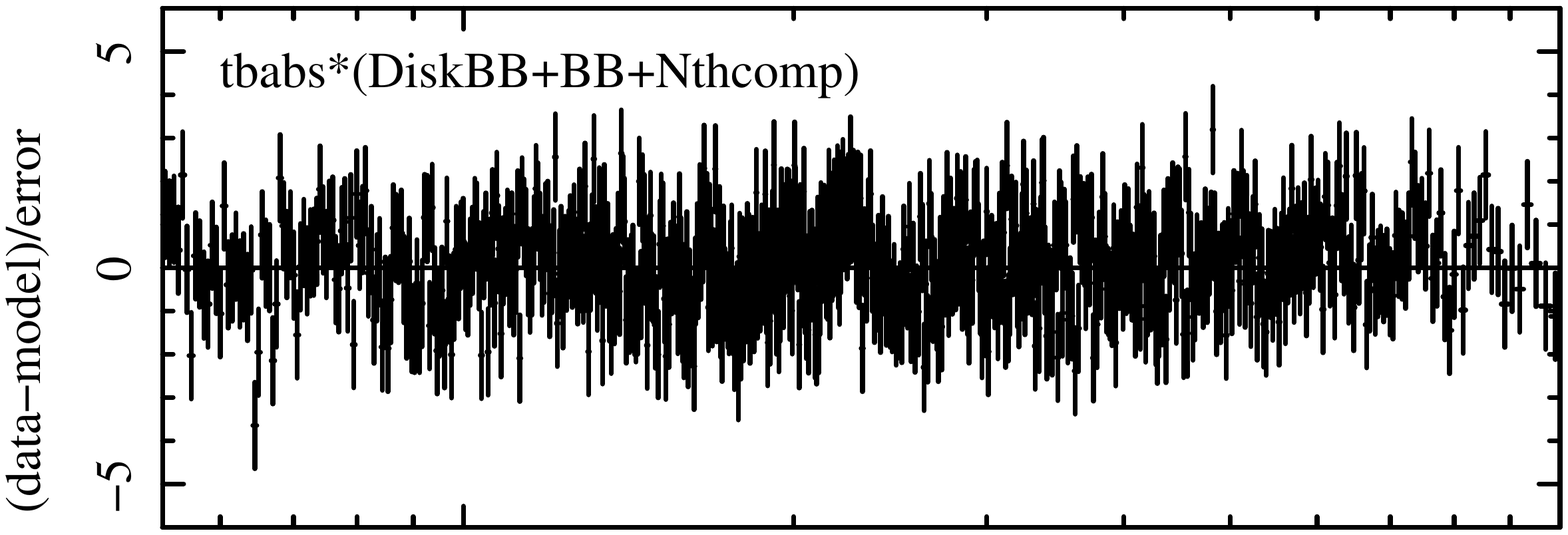}

\vspace{-4.52cm}
\includegraphics[height=0.4\textwidth,angle=0]{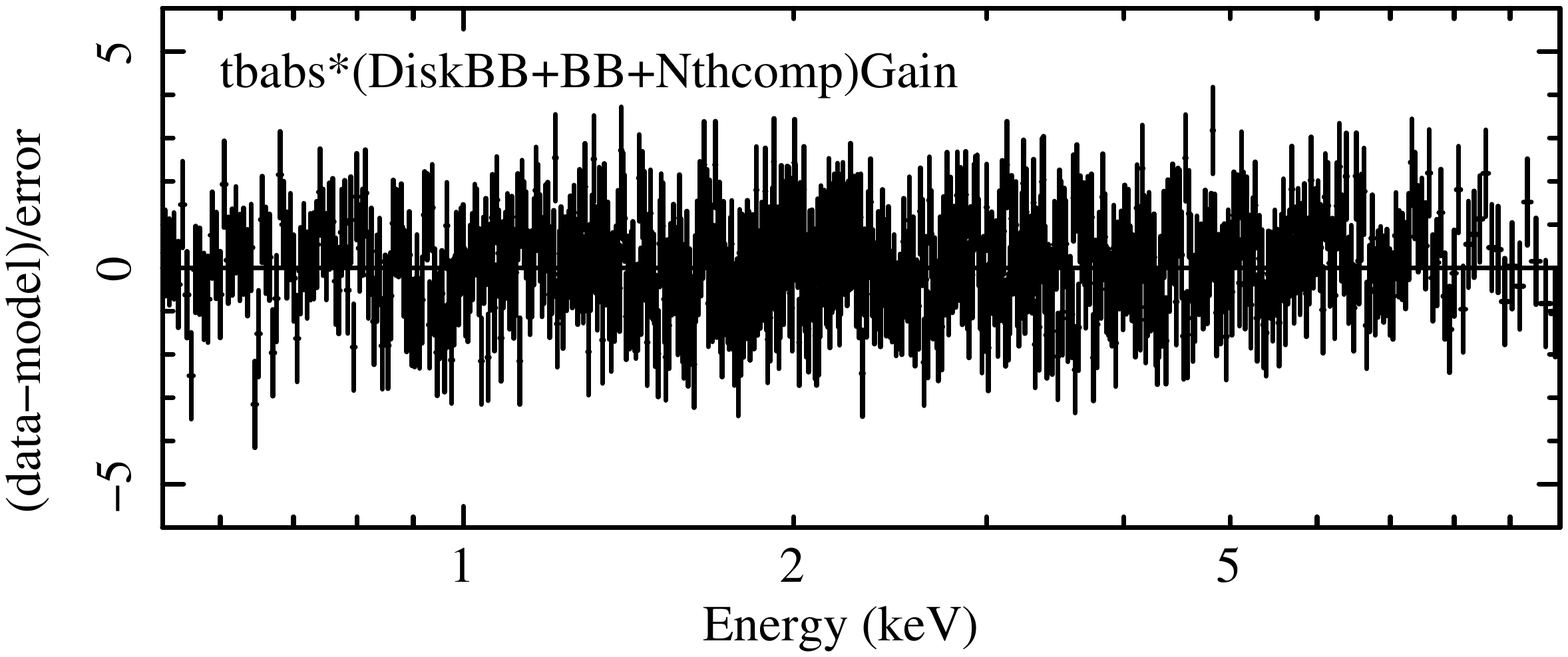}
\caption{{\it (Top panel)} \xmm\ EPIC-pn hard state spectrum fitted 
by the best fit model (disc black body plus blackbody and a dominant 
Comptonisation components absorbed by neutral material).
No signs of ionised absorption is present. 
{\it (Centre top panel)} Residuals once the data are fitted with an absorbed 
power law. 
{\it (Centre panel)} Residuals after fitting with an absorbed disk black 
body plus Comptonisation model.
{\it (Centre bottom panel)} Residuals after the addition of a black 
body component.
{\it (Bottom panel)} Residuals after the consideration of a gain 
offset in the response files. }
\label{xmmSPEC}
\end{figure} 
The black data in Fig. \ref{xmmSPEC} show 
a clear drop, in the spectrum, at energies below $\sim1$~keV, suggesting the presence of neutral 
absorbing material. We fitted the persistent spectrum with an absorbed 
power law spectrum ({\sc tbabs*powerlaw} in {\sc xspec}). 
Although this model can reasonably reproduce the bulk of the observed 
X-ray emission ($N_H=2.42\pm0.02\times10^{21}$~cm$^{-2}$, 
$\Gamma_{PL}=1.82\pm0.05$), large residuals make the fit unacceptable 
($\chi^2=2351.8$ for 1782 dof; Tab. 4). 

Therefore, we added a disc black-body component ({\sc diskbb}) 
and substituted the power law emission with a Comptonisation component 
({\sc nthcomp}) absorbed by neutral material. 
We assumed a temperature of $kT_e=17$~keV for the Comptonising 
electrons. This temperature corresponds to the best fit value of the 
fit of the \nustar\ data, accumulated two days after the \xmm\ observation 
(see Degenaar et al. in prep). This model well reproduced the data 
($\chi^2=1844.5$ for $1780$ dof). 
The spectrum was dominated by the Comptonisation component 
with $\Gamma\sim1.8$ and a cold ($kT_{DBB}=0.19\pm0.01$~keV) 
disk blackbody component, with a large inner disk 
radius ($r_{DBB}\sim150-200$~km) required by the data. 

The further addition of a blackbody emission component slightly 
improved the fit ($\chi^2=1820.9$ for 1778 dof, with associated 
F-test probability $\sim1\times10^{-5}$). The best fit blackbody 
component was hotter ($kT_{BB}=0.45\pm0.05$~keV) than the disc 
blackbody one (however colder than what is typically observed 
during the soft state) and produced by a small area with a radius 
of $\sim9$~km. 

We noted a significant excess occurring at $\sim2.2$ keV, where the effective 
area experiences a drop due to the Au M edge. 
To determine the nature of such excess, we modified\footnote{We 
performed this by employing the {\sc gain} model to the response files.  
The {\sc gain} command shifts the energies on which the response 
matrix is defined as well as the effective area in such a way that 
$E_{new}=E/sl - gain_{off}$ (where $E$ and $E_{new}$ are the original 
and shifted energies). We searched for the best fit offset in the gain, 
however we kept the slope fixed to the default value ($sl=1$).} 
the gain in the response files until reaching the best fit. 
We observed a significant improvement of the fit ($\Delta\chi^2=36.1$
for 1 dof) with a gain offset of $gain_{off}=12\pm3$~eV. Such offset is smaller 
than the uncertainty on the calibration of the EPIC energy scale (see 
XMM-SOC-CAL-TN-0018). This suggested that such residual is instrumental 
and it will not be discussed further\footnote{Please note that we did not use 
the {\sc gain} model any further in this work. }.

Based on this best fit hard state model and the constraints 
from \swift-UVOT and \nustar\ (see Degenaar et al. in prep) data, 
we reconstructed the "bona-fide" hard state SED (black line, Fig. \ref{SED}).
The observed and un-obscured 0.1-100 keV flux are 
$F_{0.1-100}=2.9\times10^{-10}$ and $3.9\times10^{-10}$
erg~cm$^{-2}$~s$^{-1}$, respectively. 
The disc black body, black body and Comptonisation components 
carry $\sim15$~\%, $\sim1$~\% and $\sim84$~\% of the 
un-obscured 0.1-100 keV flux. 
\begin{table}
\scriptsize
\begin{tabular}{ l l c c c c c}
\hline
\hline
\multicolumn{6}{c}{\bf Hard state - \xmm} \\
\hline
\hline
Model       &           & {\sc H1}       & {\sc H2}        & {\sc H3}        &{\sc H3gain}    \\
$N_{H}$     &$\star$    & $0.242\pm0.02$ & $0.31\pm0.01$   & $0.33\pm0.02$   & $0.33\pm0.01$  \\
$\Gamma$    &           & $1.82\pm0.05$  & $1.810\pm0.004$ & $1.76\pm0.02$   & $1.74\pm0.02$   \\
$N_{PL}$    & $\diamond$& $2.66\pm0.02$  &                 &                 &                 \\
$kT_{e}$    & $\flat$   &                & $17\dag$        & $17\dag$        & $17\dag$        \\
$N_{nth}$   & $\diamond$&                & $2.58\pm0.04$   & $2.4\pm0.1$     & $2.4\pm0.1$   \\
$kT_{DBB}$  & $\flat$   &                & $0.18\pm0.01$   & $0.18\pm0.01$   & $0.17\pm0.01$  \\
$N_{DBB}$   & $\ddag$   &                & $1900\pm600$    & $2400\pm1200$   & $3600\pm1200$ \\
$kT_{BB}$   & $\flat$   &                &                 & $0.45\pm0.05$   & $0.42\pm0.07$   \\
$N_{BB}$    & $\ddag$   &                &                 & $9\pm5$         & $10^{+10}_{-4}$ \\
$gain_{off}$& $\mid$    &                &                 &                 & $12\pm3$        \\
$\chi^2/dof$&           & 2351.8/1782    & 1844.5/1780     & 1820.9/1778     & 1784.8/1777    \\
\hline
\hline
\end{tabular} 
\caption{Best fit parameters of the \xmm\ EPIC-pn hard state spectrum. 
A variety of spectral components, described in the text, are employed. 
Model H1 ({\sc tbabs(powerlaw)}) is composed by a powerlaw
absorbed by neutral material ({\sc tbabs}). Model H2 
({\sc tbabs(diskbb+nthcomp)}) considers the emission from a disk 
black body plus a thermal Comptonisation component. 
Model H3 additionally considers the emission from a blackbody 
component ({\sc bbody}: {\sc tbabs(diskbb+bbody+nthcomp)}). 
Model H3gain applies a fit to the offset of the gain (with fixed slope) 
of the response matrices ({\sc gain * tbabs(nthcomp+diskbb+bbody)}). 
The following symbols mean: 
$\dag$ Fixed parameter; $\mid$ In eV units; $\diamond$ Units of $10^{-2}$. 
Other symbols as in Tab. 1. Error bars indicate 90~\% confidence, unless 
stated otherwise. }
\label{tabPer}
\end{table} 

No absorption line due to ionised iron appeared in the data. 
Indeed, we added a narrow ($\sigma=1$~eV) absorption line with a 
Gaussian profile at the energy of the \Fevc\ and \Fevs\ transitions
and we obtained upper limits to their equivalent width of 
$EW_{\Fevc}>-10$~eV and $EW_{\Fevs}>-13$~eV, respectively.  
The addition of a diskline with the same profile as observed 
during the soft state ($\alpha_d=-2.6$) did not improve the fit, 
with an upper limit on the line equivalent width of $EW<80$~eV. 

No ionised emission or absorption line are detected in the RGS
spectra, with upper limits of $\sim4-8$ eV on the Ly$\alpha$ transitions 
of the most abundant elements. We note, however, that such upper 
limits do not exclude that soft X-ray lines as observed during 
the soft state are present in the hard state.

\section{Discussion}

We studied the evolution of the X-ray emission of \mxb, throughout 
the states, during its most recent outburst,  
We observed that a three component continuum model (disk black body 
and Comptonisation in addition to thermal emission from, e.g., 
the boundary layer) provided an adequate description of the data
in both spectral states. 
The observed trends of the best fit parameters (i.e., lower disc 
temperatures, larger disc inner radii and larger Comptonisation 
fractions in the hard state) are in line with what is typically observed 
in accreting BH and-or other atoll sources (Lin et al. 2007; Dunn et al. 2010; 
Munoz-Darias et al. 2014; Armas-Padilla et al. 2018).

The normalisation of the disk black body component suggests 
that the accretion disc extends to radii comparable to the NS 
radius during the soft state, while it might be truncated 
during the hard state. In line with this, we observed broad 
residuals in the Fe~K band, in the soft state. Such residuals are well 
reproduced by a relativistic disc line model with a standard line 
equivalent width ($EW\sim250$~eV), with a disc emissivity 
$\alpha_d\sim-2.6$, off an highly inclined accretion disc ($i=75^\circ$) 
extending to few gravitational radii. Such broad feature 
appears less prominent during the \xmm\ and \nustar\ hard state 
observations. 

The persistent source emission is absorbed by a moderate 
column density of lowly ionised material that appears 
to vary from the soft ($N_H\sim0.21\pm0.01\times10^{22}$~cm$^{-2}$) 
to the hard state ($N_H\sim0.33\pm0.01\times10^{22}$~cm$^{-2}$).  
The observed variation is of the same order of magnitude 
of the one observed by Cackett et al. (2013) (i.e., from 
$N_H\sim0.2\pm0.01\times10^{22}$~cm$^{-2}$ to 
$N_H\sim0.47\pm0.13\times10^{22}$~cm$^{-2}$) during quiescence. 
Despite the origin of such variation is uncertain, it is clear 
than during outbursts dipping X-ray binaries can display variations 
of the column density of neutral absorption, by orders of magnitude 
(White \& Mason 1985; Frank et al. 1987; Ponti et al. 2016). 
Therefore, despite we analysed only the persistent emission, 
the observed variation (of $\sim40$~\%) is not surprising. 
Indeed, the observed fast variation suggests that at least 
the variable part of the column density of absorbing material 
is local to the source. If so, such material can potentially change 
its physical conditions (e.g., ionisation, location), producing 
the observed variation. 
Additionally, we note that part of such difference could 
be driven by the distorting effects of dust scattering (e.g., 
induced by combination of the energy 
dependence of the dust scattering cross section and the different 
spectral extraction regions; Tr\"umper \& Sch\"onfelder 1973; 
Predehl \& Schmitt 1995; Jin et al. 2017; 2018). 
Therefore, we caution the reader from deriving conclusions 
from this difference\footnote{In theory, by characterising 
the shape and the variability of the dust scattering halo, 
it is possible to correct the spectrum of the source, therefore 
to verify the reality of the column density variation (see 
for example the case of \axj; Jin et al. 2017; 2018). 
Despite we recently developed such spectral corrections for various 
extracting regions of modern X-ray telescopes, unfortunately 
these are not applicable to first order grating data, because 
of the complexity of the source extraction in dispersed spectra; 
Jin et al. 2017; 2018).}. 

\subsection{Ionised absorber - state connection}

\mxb\ shows clear evidence for highly ionised (Fe~K) absorption 
during the soft state that disappears (at least the \Fevc\ and \Fevs\ lines) 
during the hard state. 
This similar behaviour is shared also by two of the best monitored atoll sources 
(Ponti et al. 2014; 2015; 2017) and it is ubiquitous in accreting BH
binaries (Ponti et al. 2012)\footnote{Exceptions to this trend have been reported 
during the hard state outburst of V404~Cyg (King et al. 2015) and in some 
Z-sources (Homan et al. 2016). The latter class of sources are characterised 
by more complex behaviours and state classification as well as larger luminosities, 
compared to atoll NS.}. We note that, in BH systems, 
the observed wind - state connection has been occasionally ascribed 
to a variation of the geometry of the wind, or a variation of the magnetic field 
configuration, or a variation of the launching mechanism, etc. 
(Ueda et al. 2010; Miller et al. 2012; Neilsen et al. 2012). 

\subsection{Stability of the disc atmosphere}

As discussed in \S~\ref{stability}, the best fit ionised absorber 
parameters ($log(\xi/erg~cm~s^{-1})=3.76$; 
$log(N_H/cm^{-2})=23.48$) correspond 
to a thermally stable solution during the soft state (Fig. \ref{Stability}). 
Given the properties of the disc atmosphere observed during 
the soft state, we can predict its behaviour during the hard state, 
under the reasonable assumption that the variations of the absorber 
are solely due to the different illuminating SED. Indeed, 
as a consequence of the variation of the SED, the ionisation parameter
is expected to vary following $\xi=L/(nR^2)$, if the density ($n$) 
and location ($R$) of the absorber remain unchanged (or that they 
both vary keeping $nR^2$ constant). 
Indeed, in such a situation, the ionisation parameter of 
the atmosphere would change in response to the variations 
of the source luminosity ($L$) following the relation: 
$\xi=L/(nR^2)$. Hence, we can estimate the expected 
ionisation parameter during the hard state as: 
$\xi_h=(\xi_s L_h) / L_s$ (where $L_s$, $L_h$, $\xi_s$ and 
$\xi_h$ are the soft and hard luminosities in the 0.1-100~keV
energy band and ionisation parameters, respectively), 
corresponding to $log(\xi_h)\sim3.30$ (see Fig. \ref{Stability}). 
Therefore, during the hard state 
the ionisation parameter is lower than during the soft state. 
Indeed, despite the shape of the SED is harder, the lower 
hard state total luminosity implies lower ionisation. 
Clearly, the hard state spectrum rules out 
the presence of an absorber with 
$log(\xi_h/erg~cm~s^{-1})=3.30$ and $log(N_H/cm^{-2})=23.48$. 
This implies that the ionised absorber must have changed 
(i.e., $nR^2$ or both) as a result of the variation of the SED.
Indeed, we note that the estimated hard state parameters 
of the ionised absorbers fall onto an unstable branch of the 
thermal photo-ionisation stability curve (Fig. \ref{Stability}). 
Therefore, most likely the absorber will migrate away from 
such unstable equilibrium, 
possibly towards a stable higher and-or lower ionisation 
parameter (see Bianchi et al. 2017 for more details). 
We note that such behaviour is common to the classical 
hard (and hard-intermediate) state SED of BH LMXB. Therefore, this 
mechanism could be responsible for the observed ionised 
absorber - state connection.  

\subsection{Distance to the ionised absorber}
\label{distance}

The constraints on the ionised absorber density obtained 
in \S~\ref{doublet} from the \Fetwt\ doublet ratio can be used 
to estimate its distance from the irradiating X-ray source ($R$). 
Indeed, using the equation: $R^2=L/n\xi$, 
we derived $R<1.5\times10^{10}$~cm, for a source luminosity 
$L\sim1.3\times10^{37}$~erg~s$^{-1}$ and plasma ionisation
parameter $log(\xi/erg~cm~s^{-1})\sim3.76$, as observed during the soft state. 
Such distance corresponds to $R\leq7\times10^4$~r$_g$ and it is 
a significant fraction of the accretion disc size.  

Assuming the outer disc radius ($R_D$) to be 
$\sim80$~\% of the primary's Roche-lobe radius ($R_L1$), 
calculated by the approximate formula: 
$R_D= 0.8a \times (0.46 q^{-2/3})/[0.6 q^{-2/3} + log(1+q^{-1/3})]$, 
where $q$ is the mass ratio $m_2/m_1$ between the masses 
of the primary and companion star in Solar masses (Eggleton 1983) 
and $a$ is the binary separation which can be conveniently 
expressed in the form: 
$a=3.5\times10^{10} (1+q)^{1/3} m_1^{1/3} P_{orb}^{2/3}$~cm, 
where $P_{orb}$ is the orbital period in days (King et al. 1996), 
we computed the outer disc radius (for $P_{orb}=7.11611$~hr, 
$m_1=1.4$ and $m_2=0.6$~M$_\odot$; Ponti et al. 2018) to be 
$R_D=6.2\times10^{10}$~cm or $\sim3.0\times10^5$~r$_g$. 
Therefore, the ionised plasma sits well inside the disc. 

We also note that the plasma is located inside the 
Compton radius too. To estimate the latter, we first computed 
the Compton temperature ($T_{C}$) by integrating the observed 
soft state SED and obtaining: $T_C=1.57\times10^7$~K. 
We then derived the Compton radius ($R_C$) and the critical 
luminosity ($L_{cr}$) from the formulae: 
$R_C=(10^{18} \times m_1) / T_C$~cm and 
$L_{cr}=0.03 L_{Edd} / \sqrt{T_C/10^8}$, where $L_{Edd}$ 
is the Eddington luminosity (Begelman et al. 1983). 
The Compton radius resulted to be $R_C=8.9\times10^{10}$~cm 
and $L_{cr}=1.5\times10^{37}$~erg~s$^{-1}$. 
We note that both the ionised plasma sits well inside the 
Compton radius ($R<0.17\times R_C$) and the soft state 
luminosity ($L\sim 0.87\times L_{cr}$) is comparable if not 
smaller than the critical luminosity, therefore hampering 
the generation of a thermally driven wind. 
For these reasons, no thermal wind can be generated 
in \mxb\footnote{Although this does not apply to magnetic winds.} 
and the absorber can be at rest, forming an accretion disc 
atmosphere. 

\subsection{Single or multi-layer absorber and other approximations} 

The best fit model suggests that a single layer absorber with high 
column density and ionisation parameter can reproduce 
the bulk of the absorption features in the average spectrum. 
At first, this appears rather surprising. Indeed, the large array 
of different transitions with significantly different ionisation 
potential might carry information of a radial stratification 
of the ionisation of the absorber, or be the consequence 
of significant variations over time of the plasma parameters. 
However, before properly addressing these very interesting 
scenarios, a detailed understanding of secondary effects impacting 
the expected results need to be fully carried out. 
Indeed, several of the strongest absorption lines are saturated 
(Ponti et al. in prep), the ionised absorber is likely variable, 
the line broadening might be different from the value assumed 
in this work ($v_{turb}=500$~km~s$^{-1}$). 
We therefore leave such questions for future investigations. 

\section{Conclusions}

We studied the evolution of the X-ray emission of the accreting 
NS LMXB \mxb, comparing the soft and hard accretion states. 
During the soft state, we detected 60 absorption lines (of which 
21 at more than $3 \sigma$), that is about one order of magnitude 
more lines compared to previous works (6 lines were detected by 
Sidoli et al. 2001; Diaz Trigo et al. 2006). 
This allows us to place tighter constraints on the physical properties 
of the absorbing plasma ($log(\xi/erg cm s^{-1})\sim3.8$, 
$log(N_H/cm^{-2})\sim23.5$), to demonstrate that both the low 
energy transitions and the Fe K lines are at rest, 
without the requirement to invoke a mildly outflowing 
disc atmosphere. The repeated detection of an high ionisation 
absorber during the soft states in two different outbursts (Sidoli et al. 
2001; Diaz Trigo et al. 2006; this work) suggests the persistent presence 
of such component during the soft state. 
The new data show that the high ionisation absorption (traced by 
the \Fevc\ and \Fevs\ lines) significantly weakens/disappears 
during the hard state, indicating a physical change of the 
absorber (disc atmosphere) between the two states. 
This is likely a consequence of a photo-ionisation thermal instability 
that drives the Fe K absorber to new equilibria consistent 
with fully or lowly ionised plasma (Bianchi et al. 2017). 
Thanks to the tentative detection of one of the 
lines of the \Fetwt\ doublet, we also constrained the density 
of the absorber ($n_e>10^{13}$~cm$^{-3}$).

\subsection*{Comparison with similar works}

{\it While this work was refereed, a similar paper has been 
accepted (Sharma et al. 2018) and another has been submitted and 
posted on arXiv (Iaria et al. subm.). Sharma et al. (2018) finds very 
similar results to the ones presented here. Different components are, 
instead, investigated by Iaria et al. However, we refrain from performing 
a detailed comparison, because that work is still in the refereeing 
process. }

\section*{Acknowledgments}

The authors wish to thank Norbert Schartel and Fiona Harrison 
for approving the \xmm\ and \nustar\ DDT observations of \mxb. 
GP acknowledges financial support from the Bundesministerium 
f\"{u}r Wirtschaft und Technologie/Deutsches Zentrum f\"{u}r Luft- 
und Raumfahrt (BMWI/DLR, FKZ 50 OR 1812, FKZ 50 OR 1715 
and FKZ 50 OR 1604) and the Max Planck Society. 
SB acknowledges financial support 
from the Italian Space Agency under grants ASI-INAF I/037/12/0 
and 2017-14-H.O. BDM acknowledges support from the Polish 
National Science Center grant Polonez 2016/21/P/ST9/04025 
and from the European UnionÕs Horizon 2020 research 
and innovation programme under the Marie Sklodowska-Curie 
grant agreement No 798726.
ND is supported by a Vidi grant from the Netherlands 
Organization for Scientific research (NWO).

\end{document}